\documentclass{aa}

\usepackage{amsmath,amssymb,amsthm}
\usepackage{mathabx,mathrsfs}
\usepackage{multicol}
\usepackage{natbib}
\usepackage{graphicx}
\usepackage{svg}
\usepackage{txfonts}
\usepackage{atbegshi}
\usepackage[para,online,flushleft]{threeparttable}
\bibpunct{(}{)}{;}{a}{}{,}

\newcommand{\orcid}[1]{\unskip\protect\href{https://orcid.org/#1}{\protect\includegraphics[width=8pt,clip]{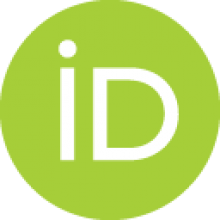}}}

\newcommand{\kpc}{{\rm kpc}}
\newcommand{\mh}{{\rm [M/H]} }

\newcommand{\am}{[$\alpha$/M]}
\newcommand{\dex}{{\rm ~dex}}
\newcommand{\teff}{$T_{\rm eff}$}
\newcommand{\logg}{$\log~g$}
\defcitealias{spitoni_2021}{S21}
\usepackage{hyperref}
\begin{document}

    \title{
    Reconstructing the Milky Way chemical map with Galactic Chemical Evolution tool \texttt{OMEGA+} from SDSS-MWM 
    }
    \author{Viola~Heged\H{u}s\inst{1,2,3}\orcid{0000-0001-7699-1902}
    \and 
    Szabolcs~M{\'e}sz{\'a}ros\inst{1,2,3}\orcid{0000-0001-8237-5209}
    \and
    Blanka~Vil{\'a}gos\inst{4,5}\orcid{0000-0001-9688-2332}
    \and
    Marco~Pignatari\inst{4,6,7,8}\orcid{0000-0002-9048-6010}
    \and 
    Emily J. Griffith\inst{9, 10}\orcid{https://orcid.org/0000-0001-9345-9977}
    \and 
    Diogo Souto\inst{11}\orcid{https://orcid.org/0000-0002-7883-5425}
    \and
    Maria~Lugaro\inst{4,6,3, 12}\orcid{0000-0002-6972-3958}    
   }

\institute{ELTE E\"otv\"os Lor\'and University, Gothard Astrophysical Observatory, 9700 Szombathely, Szent Imre H. st. 112, Hungary    
    \and    
    ELTE Eötvös Loránd University, Institute of Physics and Astronomy, Budapest 1117, Pázmány Péter sétány 1/A, Hungary
    \and
    MTA-ELTE Lend{\"u}let "Momentum" Milky Way Research Group, Hungary
    \and
    Konkoly Observatory, HUN-REN CSFK/RCAES, Konkoly Thege Miklós út 15-17, H-1121 Budapest, Hungary
    \and
    The Oskar Klein Centre, Department of Astronomy, Stockholm University, AlbaNova, SE-10691 Stockholm, Sweden
    \and
    CSFK HUN-REN, MTA Centre of Excellence, Budapest, Konkoly Thege Miklós út 15-17., H-1121, Hungary
    \and
    E. A. Milne Centre for Astrophysics, University of Hull, Cottingham Road, Kingston upon Hull, HU6 7RX
    \and
    NuGrid Collaboration, \url{http://nugridstars.org}
    \and 
    Center for Astrophysics and Space Astronomy, Department of Astrophysical and Planetary Sciences, University of Colorado, 389 UCB, Boulder, CO 80309-0389, USA
    \and
    NSF Astronomy and Astrophysics Postdoctoral Fellow
    \and
    Departamento de F\'isica, Universidade Federal de Sergipe, Av. Marcelo Deda Chagas, S/N, 49107-230 S\~ao Crist\'ov\~ao, SE, Brazil
    \and 
    School of Physics and Astronomy, Monash University, VIC 3800, Australia
}

\date{Submitted ...; accepted ...}

\abstract
{Although current observations show that there are two distinct sequences of disk stars in the [$\alpha$/M] versus [M/H] parameter space, there is further complexity in the chemical makeup of the Milky Way, suggesting a complicated evolutionary history.}
{We develop two-infall galactic chemical evolution (GCE) models consistent with the Galactic chemical map.}
{We obtain new GCE models simulating the chemical evolution of the Milky Way as constrained by a golden sample of 394,000 stellar abundances of the Milky Way Mapper survey from the 19th data release of SDSS-V. The separation between the chemical thin and thick disks is defined using [Mg/M]. We use the chemical evolution environment \texttt{OMEGA+}, combined with Levenberg-Marquardt and bootstrapping algorithms for optimization and error estimation. We simulate the entire Galactic disk and consider six galactocentric regions, allowing for a more detailed analysis of the formation of the inner, middle, and outer Galaxy. We investigate the evolution of $\alpha$, odd-Z, and iron-peak elements: 15 species altogether.}
{The chemical thin and thick disks are separated by Mg observations, which the other $\alpha$-elements show similar trends with, while odd-Z species demonstrate different patterns as functions of metallicity. In the inward Galactic disk regions the locus of the low-Mg sequence is gradually shifted toward higher metallicity, while the high-Mg phase is less populated. The best-fit GCE models show a well-defined peak in the rate of the infalling matter as a function of the Galactic age, confirming a merger event about 10~Gyr ago. We show that the time-scale of gas accretion, the exact time of the second infall as well as the ratio between the surface mass densities associated to the second infall event and the formation event vary with the distance from the Galactic center. According to the models, the disk was assembled within a timescale of (0.32$\pm$0.02)~Gyr during a primary formation phase, then a (0.55$\pm$0.06)~Gyr-timescale, increasing accretion rate was followed by a relaxation that lasted (2.86$\pm$0.70)~Gyr, with a second peak of the infall rate at (4.13$\pm$0.19)~Gyr.}
{Our best Galaxy evolution models are consistent with an inside-out formation scenario of the Milky Way disk, in agreement with the findings of recent chemo-dynamical simulations.}


\titlerunning{Modelling the GCE with \texttt{OMEGA+}}
\authorrunning{Heged\H{u}s et al. 2025}
\maketitle

\section{Introduction}\label{intro}

Milky Way (MW) stars of the Galactic plane form a dichotomy in the metallicity versus elemental abundance ratio parameter space, as first discussed by \citet{mcwilliam_1997}, and also by \citet{fuhrmann_1998}. The $\alpha$-elements such as O, Mg, S, Si, Ca, Ti are mainly produced by massive stars and therefore carry essential importance in distinguishing two populations of stars. The so-called high-$\alpha$ (rich in $\alpha$-elements) sequence forms the thick disk \citep{gilmore_1983}, while the thin disk consists of the low-$\alpha$ stellar population. These are often referred to as “low-Ia” and “high-Ia”, respectively, because thin disk stars are generally richer in iron, produced primarily by SNe type Ia. These sequences are also separated kinematically \citep{gaia_2018}, and \citet{silva_2018} showed that the two sequences are characterized by two different ages: the high-$\alpha$ and low-$\alpha$ sequences peak at $\sim$11~Gyr and $\sim$2~Gyr.

A review of the recent semi-numerical Galactic Chemical Evolution (GCE) model results for the MW can be found in \citet{matteucci_2021}. 
The first comprehensive analytical model assumed the instantaneous recycling approximation, which neglects the stellar lifetimes above $1~ M_{\odot}$ \citep{tinsley_1980T}. Other pioneering analytical studies \citet{pagel_1995} and \citet{pagel_1997} investigate the evolution of primary elements in the Galactic disk and in the solar neighborhood. 
There are many approaches to GCE modeling including serial, parallel, stochastic, and stellar accretion approaches, which we summarize here. 
First, the halo, thick and thin disks were considered to have formed after one another during one continuous gas infall event \citep[e.g., ][]{chiosi_1980, boissier_1999}. Studies extending this serial approach assumed instead two, independent but sequential accretion episodes \citep{chiappini_1997}. This classical two-infall model assumes that the halo and thick disk formed during the first peak of infall rate, while the thin disk assembled on a longer timescale, reproducing the abundance patterns of the thick and thin disk stars, and a gap in the SFR between the two phases sequentially. In such models, we define $t_{\rm max}$ as the time for the maximum infall onto the thin disk, and $\tau_1$ and $\tau_2$ as the characteristic timescales for gas accretion during the formation of the halo-thick disk and the thin disk.

According to the parallel approach, the gas accretion starts at the same time and occurs in parallel, but at different rates for each infall episode \citep[e.g., ][]{pardi_1995, grisoni_2017}. 
The stochastic approach was motivated by the large scatter of the chemical abundances of neutron-capture elements for halo stars at low metallicities ([Fe/H] $\lesssim -3$~dex). This scatter reflects that the pollution by a single SNe was not efficiently mixed. The stochastic approach is often applied to the stellar halo \citep[e.g., ][]{cescutti_2008}. 
Finally, the assumption of the stellar accretion means that the Galactic halo accreted stars from the dwarf satellite galaxies of the MW \citet[e.g., ][]{prantzos_2008}. This concept was presented by \citet{searle_1978}, who suggested that the building blocks of the outer halo hierarchically originated from fragments of (dwarf) galaxies over a long timescale. 

The inside-out mechanism means that the disk forms by gas accretion much faster in the inner than in the outer disk regions, thus creating a gradient in the SFR and the [$\alpha$/M] ratios \citep{larson_1976, cole_2000, bergemann_2014}. This approach of formation has been widely adopted to reproduce the observed abundance gradients within the MW \citep[e.g.,][]{spitoni_2015, palla_2020}.
For example, in their serial approach, \citet{chiappini_1997} concluded on $t_{\rm max}=1$~Gyr for an early second infall, whereas a $\tau_1=0.8$~Gyr for the halo+thick disk, and a significantly longer $\tau_2=7-8$~Gyr thin disk assemble time at the solar vicinity. 

\citet{spitoni_2019} suggested a revised two-infall model, with the second infall occurring with a delay of $\sim 4.3$~Gyr relative to the formation, to reproduce the observed bimodality and the stellar ages. The gap in star formation predicts a loop in the model curves describing the abundance ratios in the solar vicinity (see later in e.g., Fig.~\ref{global_fit}), and a decrease in $t_{\rm max}$ produces a loop starting at lower metallicities \citep{spitoni_2019}. 
The $t_{\rm max}$ delay was also analyzed by \citet{vincenzo_2019}, proposing that a massive satellite called Gaia-Sausage/Enceladus \citep{helmi_2018} was accreted by the Galaxy during a major merger event. This scenario of a high-impact merger is in agreement with \citet{chaplin_2020} who showed that the metal-poor, high-$\alpha$ star $\nu$~Indi was originally a member of the halo, which made it possible to infer that the earliest time when the merger could have begun was 11.6~Gyr ago. 

Migration has been proposed as a complementary to two-infall \citep[e.g., ][]{spitoni_2015} or alternative \citep[e.g., ][]{buck_2020, sharma_2020, pranztos_2023} solution to explain the low-$\alpha$ and the high-$\alpha$ sequences. It has been shown by \citet{roskar_2008} that resonant scattering at corotation may migrate stars outward while the gas moves inward. 
The study from \citet{vincenzo_2020} showed that stellar migration involves old metal-rich stars, and occurs more outward than inward. However, \citet{khoperskov_2020} concluded that it has a negligible effect on the [$\alpha$/M] vs. [M/H] relation.

Chemical enrichment in galaxies is traced by the chemical elements produced by stars of different initial masses, when part of their mass is redistributed to the ISM when they die. Chemical elements have different contributions from exploding massive stars, neutron star mergers, SNe type Ia, or dying low-mass stars \citep{kobayashi_2020}. The evolution of elemental ratios is also driven by phenomena such as galactic winds, star formation history (SFH), and gas inflows and outflows. For such GCE studies, obtaining significant high-resolution spectroscopic data of stellar atmospheres is crucial. 

The Sloan Digital Sky Survey is a ground-based panoptic program now operating in its fifth phase (Kollmeier et al. 2025, in prep.), that is performing multi-epoch optical and infrared observations across the entire sky \citep{gunn_2006}. One of its projects, the Milky Way Mapper survey (MWM, Kollmeier et al. 2025, in prep.) is the successor of Apache Point Observatory Galactic Evolution Experiment \citep[APOGEE, ][]{majewski_2017}. It will obtain high-precision spectroscopic data of 5 million objects throughout the sky by 2027 to provide a dense and contiguous stellar map while mainly focusing on low Galactic latitudes. The survey employs both optical and near-infrared spectroscopy performed with BOSS \citep{smee_2013} low-resolution ($R\sim2000$) and APOGEE \citep{wilson_2019,bowen_1973} high-resolution ($R\sim22,500$) instruments, respectively. Stellar parameters and atmospheric chemical composition of each star involved in the program are derived from the high signal-to-noise (S/N) spectra. 

In this article, we perform two-infall GCE models intending to reproduce the chemical maps of the MW drawn by private data of SDSS-V MWM, published in its first data release (DR19, Meszaros et al. 2025, under rev., hereinafter M25). This study covers the evolution of $\alpha$, odd-Z and iron-peak elements, 15 species altogether. By dividing the Galaxy into six regions, we analyze the solar vicinity, the inner and outer regions of the thick-and-thin disk system. The motivation raised from the work of \citet{spitoni_2021} (hereinafter S21) following up by their prior studies \citep{spitoni_2017,spitoni_2019}, but spatially dividing the MW into twice as many galactocentric regions. 
We use the One-zone Model for the Evolution of GAlaxies \citep[\texttt{OMEGA}, ][]{cote_2017} and its extension the Two-zone Model for the Evolution of Galaxies \citep[\texttt{OMEGA+}, ][]{cote_2018}.
The OMEGA+ code is designed to run GCE simulations\footnote{\texttt{OMEGA} and \texttt{OMEGA+} are part of the NuGrid chemical evolution package and are publicly available online at \url{http://nugrid.github.io/NuPyCEE} and \url{https://github.com/becot85/JINAPyCEE}.}. For previous GCE studies using these tools we refer to \citet{pignatari_2023} for the GCE of the solar neighborhood, \citet{liang_2023} for assessing stellar yields, \citet{cote_2019} for probing the origin of r-process elements, or dwarfs galaxies \citep{cote_2017}.

\begin{figure}
\centering
\includegraphics[width=.5\textwidth]{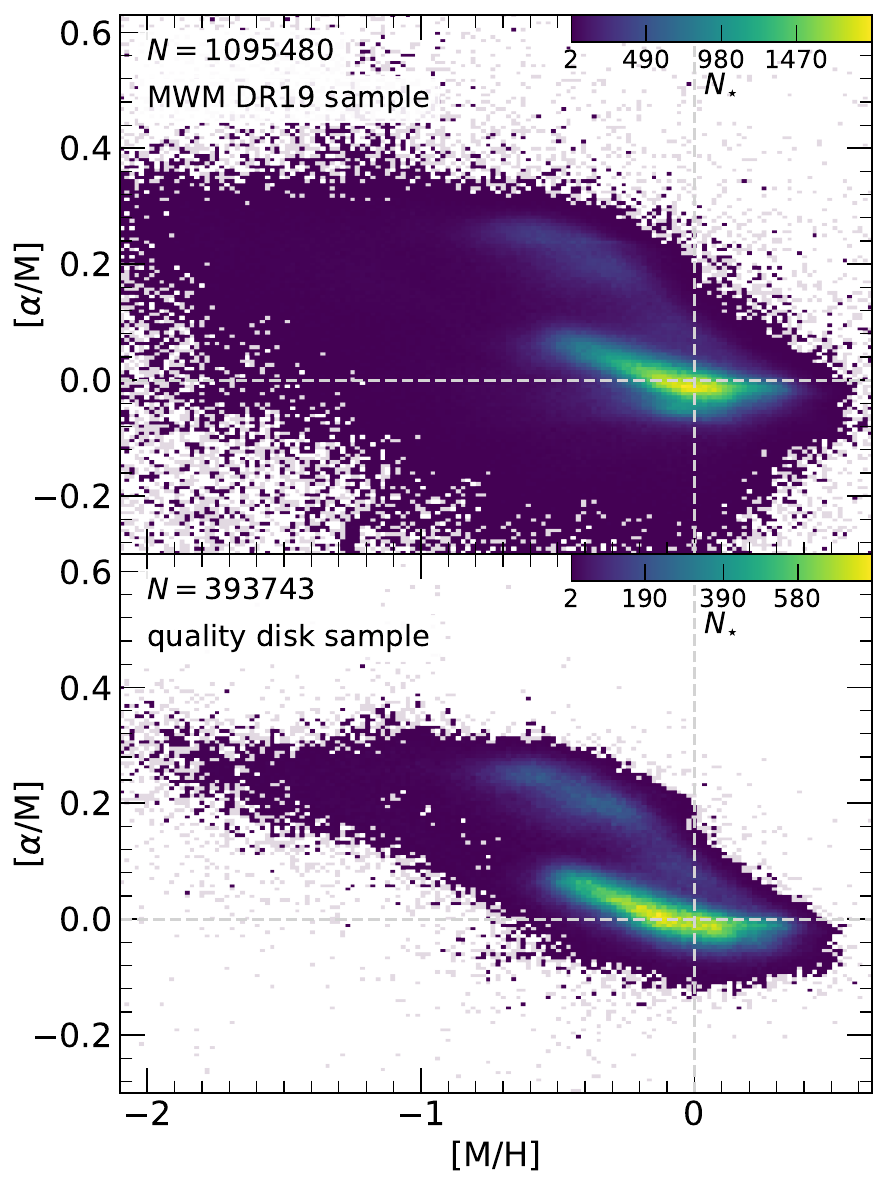}
\caption{Observed stellar [$\alpha$/M] vs. [M/H] abundance ratios from MWM DR19 (Meszaros et al. 2025, in prep.). The top panel shows a density plot for all the stars published, while the disk stars involved in the quality disk sample at the galactocentric regions $3~\kpc \leq R \leq 15~\kpc$ are depicted in the bottom panel. Color-coding of the bins starts from two, and gray squares represent bins containing a single star. The total number of stars is denoted by $N$. }
\label{datasets}
\end{figure}

\begin{figure*}
\centering
\includegraphics[width=.95\textwidth]{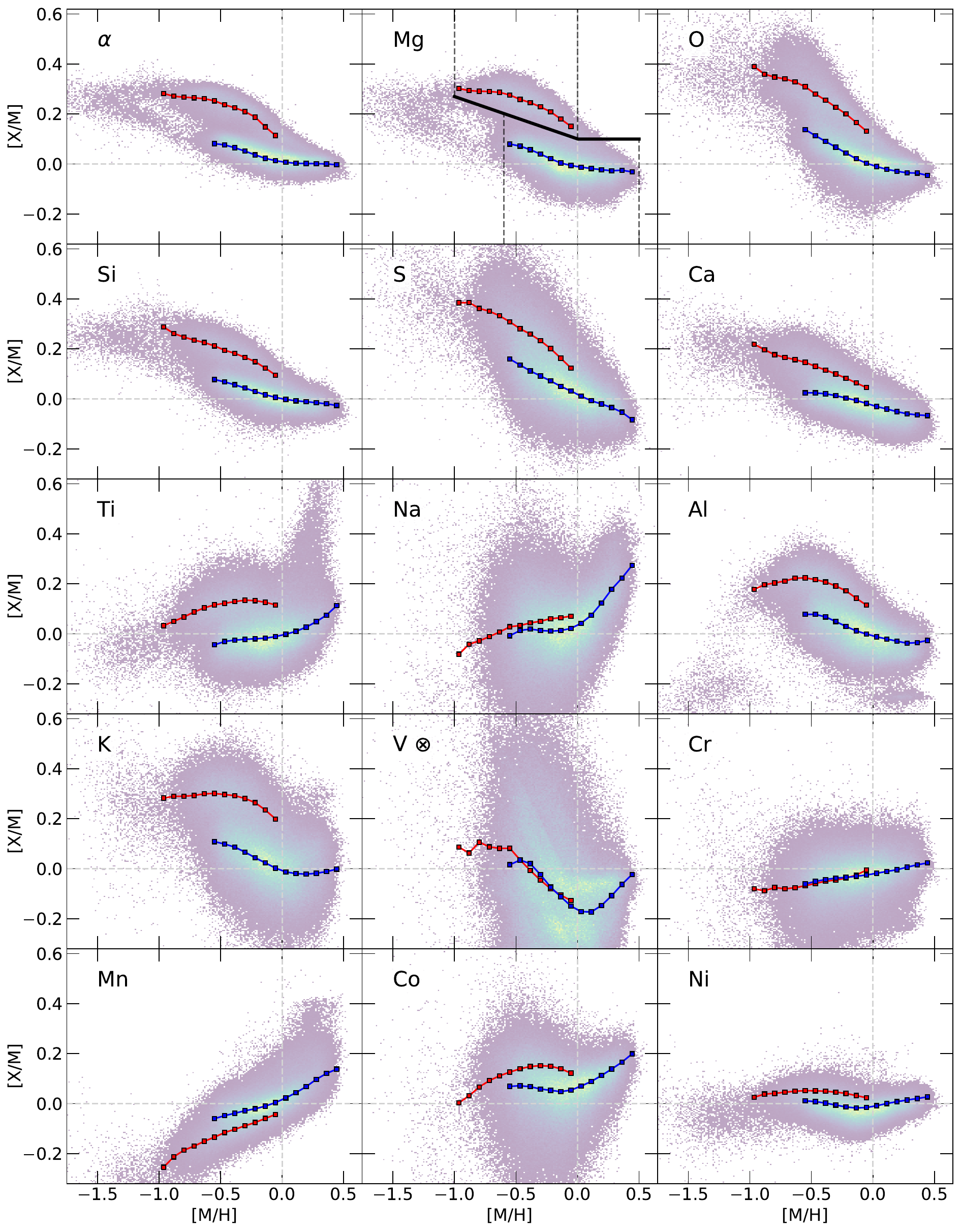}
\caption{Chemical map of the Galaxy observed by MWM. The [X/M] vs. [M/H] relations are shown for the combined $\alpha$-elemental ratio as well as the single elements from Mg to Ni for the global disk sample. Red and blue curves represent the median trends of the high- and low-Mg stars  separated based on their Mg abundance. This separation and the metallicity boundaries of the thick and thin disk stars are indicated by the black solid and dashed lines in the panel of Mg. Note that $\otimes$ drives caution to V, because of high observational uncertainty.}
\label{chem_map}
\end{figure*}

\begin{figure}
\centering
\includegraphics[width=.41\textwidth]{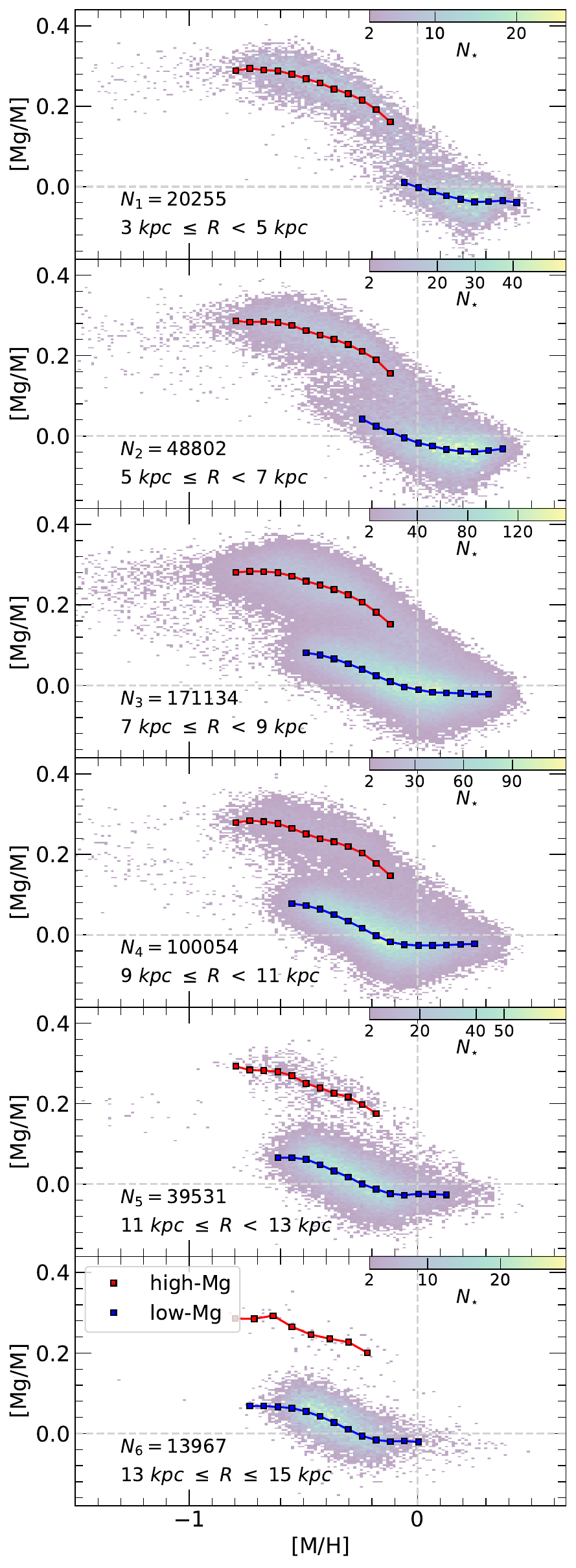}
\caption{Observed stellar [Mg/M] vs. [M/H] abundance ratios from MWM DR19 (Meszaros et al. 2025, in prep.) for the six bins of different Galactocentric distances considered here. The regions plotted from the top are all 2-kpc-wide, and are centered at $R_{1} = 4~\kpc$, $R_{2} = 6~\kpc$, $R_{3} = 8~\kpc$, $R_{4} = 10~\kpc$, $R_{5} = 12~\kpc$, $R_{6} = 14~\kpc$, respectively. The sample size in the $i$-th region is denoted by $N_{i}$. Red and blue squares represent the binned distributions of the high-Mg and low-Mg sequences, respectively. 
Details on the data selection are reported in the text.}
\label{regs_only_mg}
\end{figure}

\section{Target and data selection}\label{target_data}

\subsection{The Milky Way Mapper DR19 sample}\label{mwm}

As part of SDSS-V, one of the innovations of MWM is its new automated pipeline called \texttt{Astra} (Casey et al. 2025, in prep.) that derives stellar atmospheric parameters for the new MWM and also for the old APOGEE observations by reanalyzing those measurements. \texttt{Astra} is capable of running multiple algorithms developed to derive abundances, including the APOGEE Stellar Parameters and Chemical Abundance Pipeline  \citep[ASPCAP;][]{garcia_2016}, which relies on the FERRE\footnote{\url{github.com/callendeprieto/ferre}} multidimensional $\chi^2$ optimization code \citep{allende_2006}. Individual abundances of elements that are used in this paper are derived in the second phase of ASPCAP by fixing the main atmospheric parameters obtained in the first phase while fitting the wavelength windows sensitive to only the particular element. Note that NLTE effects were not taken into account in the spectral grid in DR19 (M25).

The first data release of MWM is publicly available online\footnote{The MWM DR19 database can be downloaded from the SDSS-V Science Archive Server: \url{data.sdss.org/sas/dr19/} .}, and contains spectroscopic, photometric and astrometric information of 1,095,480 individual stars, from which we extracted the effective temperature (\teff), surface gravity ($\log~g$), metallicity ([M/H]), calibrated abundance of the relevant element X ([X/H], where X=\{$\alpha$, O, Mg, Si, S, Ca, Ti, Na, Al, K, V, Cr, Mn, Co, Ni\}), equatorial coordinates (RA, Dec) along with parallax ($p$), as well as the quality flags. Note that the $\alpha-$capture elements considered in ASPCAP are the following: O, Mg, Si, S, Ca, and Ti. Practically, the [$\alpha$/M] abundance value reflects the O, Mg, and Si because these have the strongest lines, while S, Ca, and Ti have lower weights within the fitting of the $\alpha$-dimension (M25). The oxygen abundance is derived from OH molecular lines in the H-band, that is sensitive to \teff, although reliable between $3500$ and $6000$~K (M25). The atmospheric parameters \teff~ and \logg~ are calibrated to the photometric scale based on the infrared flux method and to asteroseismic surface gravities, respectively. For the individual abundances, zero-point offsets are applied using the solar neighborhood and solar metallicity red giant branch (RGB) stars. These calibrations and an exhaustive assessment of the overall uncertainties are explained in detail by M25. The reported values of the overall precision of \teff, \logg, metallicity and individual chemical abundance ratios for the giants are 50$-$70~K, 0.05$-$0.33~dex, 0.03~dex and below 0.15~dex for most of the elements (except for e.g., vanadium), respectively.

We use the customary scale of logarithmic abundances defined as $A(X)=\log(N_X/N_H) + A(H)$, where $A(H)\equiv12$ and $N_X$ is the number of atoms of element $X$ per unit volume in the atmosphere. The abundance ratio is defined relative to the chemical composition of the Sun: 
\begin{equation}\label{eq:abund_def}
    {\rm [X/H]} = \log \left(  {n_{X}}/{n_{H}} \right)_{{ \rm \star}} - \log \left(  {n_{X}}/{n_{H}} \right)_{\odot}.
\end{equation}
Elemental abundances are usually determined relative either to hydrogen [X/H] or iron [X/M], and the conversion between them is given by 
\begin{equation*}
    \rm [X/H] = [X/M] + [M/H].
\end{equation*}
Both the overall metallicity [M/H] and the iron abundance [Fe/H] are published in DR19 (M25). While [M/H] is determined by fitting the entire H-band spectrum as an overall scaling factor of all the metal abundances with a solar abundance pattern, [Fe/H] is derived in the second stage of the fit in wavelength windows that are sensitive to Fe~I lines. The two metallicity indicators generally provide the same values within the expected uncertainties (M25). The median difference between iron and all metal abundances is 0.01~dex with a scatter of only 0.03~dex. We refer to \mh as metallicity throughout this study.

\subsection{Target selection}\label{sec:target}

\subsubsection{Quality cuts}

This section discusses our selection criteria to obtain a reliable parameter space covering the Galactic disk. We opted out stars that are marked with the quality control \texttt{flag\_bad} by the ASPCAP pipeline, as well as if one of the atmospheric parameters (\teff, \logg, [M/H], [Mg/M], [$\alpha$/M]) is missing. Moreover, we applied a S/N cut of 50 so that we could use the most reliable parameter space proposed by MWM (M25). 
As recommended by M25, we also eliminated stars having $T_{\rm eff} < 4250$~K and [C/M] $>$ 0.1 dex. The reason is that this carbon-rich cool group cannot be accurately fitted by ASPCAP, regardless of their surface gravity. The [C/M] abundance affects all spectral regions as it is determined from the global fit of the spectra and kept fixed in the stages of determining the individual abundances.

\subsubsection{Atmospheric parameter selection}

For the main atmospheric parameters, we followed the cuts suggested by \citet{hayden_2015}, therefore only stars with \teff~$\in[3500~{\rm K};5500~{\rm K}]$, and $\log~g\in[1.0~{\rm dex};3.8~{\rm dex}]$ were retained in the sample. This $T_{\rm eff}$--$\log~g$ range contains RGB and red clump stars within the most reliable parameter range of MWM, excluding main sequence stars, that have less reliable abundances than giants in DR19 (M25).

\subsubsection{Spatial selection}

We attempt to focus on the evolution of the Galactic disk, therefore selected stars from this region of the MW according to their $R$ Galactocentric distances and $Z$ vertical heights. To perform the conversion from the (RA, Dec) equatorial coordinates and the Gaia DR3 parallax $p$ published\footnote{The published (RA, Dec, $p$) data are measured within the ICRS coordinate-system with an epoch of J2000.0.} in MWM DR19 to the ($R$, $\phi$, $Z$) cylindrical galactocentric frame, we used the \texttt{astropy} package \citep{astropy_2022}. This coordinate transformation assumes the Galactocentric distance from the Sun to the Galactic center $R_{\odot} = 8.122 ~\kpc$ and its location above the Galactic mid-plane $Z_{\odot} =20.8~{\rm pc}$ \citep{gravity_2019}. Similarly to, e.g., \citet{griffith_2021}, the selected stars thus have distances from the center and from the Galactic mid-plane of $3~\kpc \leq R \leq 15 ~\kpc$ and $|Z|< 5 ~\kpc$, respectively. This is an extended range of the vertical distance of $|Z|< 2 ~\kpc$ used by \cite{hayden_2015}.

This study does not consider the innermost central region of $R < 3$~kpc as the Galactic bulge characteristics are dominant there, and a different GCE setup is required for this central component. However, recent works \citep[e.g., ][]{queiroz_2020} based on APOGEE DR16 data confirm that a disk-like, bimodal [$\alpha$/M] vs [M/H] abundance distribution is also observed in the Galactic bulge. It was reported in \citet{queiroz_2020} that contamination caused by varying the definition for the spatial coverage of the bulge, within a range of $2-6~\kpc$, does not account for significant changes in the observed bimodal disk-like distribution. Here we confirm that MWM DR19 stars with Galactocentric distances selected from the range $R=2-6 ~\kpc$ represent an almost identical distribution in the [Mg/M] vs. [M/H] abundance ratio space as those enclosed in a region of $R=3-6 ~\kpc$. 

In addition, stars with a high probability of being globular cluster members were also removed from the sample. Our source to perform this identification was the SDSS-IV/APOGEE DR17 \citep{abdurro_2021} value-added catalog (VAC) of Galactic Globular Cluster stars, which contains membership information for 7,732 observations (6,422 unique stars) in 72 globular clusters in the MW \citep{schiavon_2024}. 
To achieve completeness in spatial filtering, we deleted stars still marked as likely globular cluster or dwarf galaxy (such as $\omega$Cen) members. Under the label of \texttt{sdss4\_apogee\_member\_flags}, a bit is set if a given target meets several membership criteria\footnote{\url{https://www.sdss4.org/dr17/irspec/apogee-bitmasks/}}. In the VAC, under the relevant flag label, candidate members are assigned based on sky position, proper motion, and radial velocity.

The \mh vs. \am~ distribution for the Galaxy of the original DR19 data set and the sample after our cuts are shown in Fig.~\ref{datasets}. Note that the separation of the low-Mg and high-Mg populations became sharper and less scattered after applying the cuts described above, and artifacts/synthetic nodulations have disappeared. While the original DR19 data set contains 1,095,480 stars, we retain 393,743 objects to use in the GCE models after applying our selection.

\subsection{Data description} \label{mod_descr}

In this work, we analyze the Galactic disk between $R = 3$~kpc and 15~kpc, dividing the sample into the concentric annular rings in steps of $\Delta R = 2~\kpc$. 

After applying the selection criteria described in Sect.~\ref{sec:target}, we create chemical maps of the Galactic disk for the total $\alpha$-elemental abundance, and also for the other 14 single elemental species: Mg, O, Si, S, Ca, Ti, Na, Al, K, V, Cr, Mn, Co, and Ni. The separation between the populations is determined by two linear functions on the Mg abundance distribution of the MW. Note that \citet{hayden_2015} and \citet{weinberg_2019} performed a similar division on the same [Mg/M] vs. [M/H] parameter plane. The boundary line is a decreasing linear function at metallicities below zero and a constant above zero, as described by
\begin{equation}\label{eq:division}
    {\rm [Mg/M]}=
    \begin{cases}
        -0.17 \cdot {\rm [M/H]} + 0.12 {\rm ~dex},  & {\rm [M/H]} \leq 0\\
        0.12 {\rm ~dex},    & {\rm [M/H]} > 0~.
    \end{cases}
\end{equation}
We determine the slope of the line regression by first dividing the distribution into 30 metallicity bins between $-1\dex<$ \mh $< 0\dex$, followed by fitting the Mg distribution in each \mh \ bin (vertical projection in the parameter-plane) with the sum of two Gaussians, then we fitted a linear function on the minimum values of these 2-peak functions. The resulting linear function is denoted by a solid line in the Mg panel of Fig.~\ref{chem_map}. 
The high-Mg sequence is defined between $-1.0\dex<\mh < 0.0\dex$, whereas the low-Mg phase starts at $\mh = -0.6\dex$ and ranges up to the metallicity of $\mh = 0.5\dex$. Outside these boundaries, the [Mg/M] distributions do not display any relevance to the sequences and/or contain only few observations. 

After performing the separation according to Eq. (\ref{eq:division}), the median values of the resulting two sets of stars were considered as the locus of the high-Mg and low-Mg populations. The thin and thick disks are also well separated by the other elements in Fig.~\ref{chem_map}, that shows the characteristic sequences defined by Mg. While most elements show similar trends to Mg, the abundance trends of odd-Z elements such as Co, Mn, and V are noticeably discrepant, showing a completely different pattern as function of metallicity. We note that Ti and V abundances derived from APOGEE data are known to be challenging to measure precisely in red giant stars, primarily due to the weakness of their spectral lines, as discussed by \citet{souto_2016,souto_2018}.

Hereinafter, the global sample refers to all stars spanning the range of $3~\kpc \leq R\leq 15~\kpc$, while the six 2-kpc-wide, distinct regions make up the inner zone (1st and 2nd regions covering $3~\kpc \leq R < 7~\kpc$), the middle zone (3rd and 4th regions covering $7~\kpc \leq R < 11~\kpc$) and the outer zone (5th and 6th regions covering $11~\kpc \leq R \leq 15~\kpc$). For the exact numbers of the subsets, see Fig.~\ref{regs_only_mg} and Table~\ref{sum_results}. The highest number of observed stars appears in the 3rd region, which includes the solar neighborhood, while the 6th region contains the least measurements  
The chemical maps in the concentric annuluses are shown in Fig.~\ref{regs_only_mg}. The two populations visible in the [Mg/M] vs. \mh \ relation have variable trends with the radius, as reported by \citet{hayden_2015} and \citet{queiroz_2020} from the APOGEE data. 
The average [Mg/M] values of the thick disk remain roughly the same between the regions, therefore the expression of Eq.~(\ref{eq:division}) used to separate the thin and thick disks can be applied for all regions uniformly.

The low-Mg population shows lower and upper boundaries at increasingly lower metallicity toward the outer disk, while the metallicity range of the high-Mg population covers the $-0.85~ {\rm dex} < {\rm [M/H]} < -0.10~ {\rm dex}$ range across all regions. In other words, the locus of the low-Mg sequence is shifted toward higher metallicity in the inner regions. It is also clear from Fig.~\ref{regs_only_mg} that stars at the outer edge of the Galaxy (bottom panel, 6th region) preferentially populate the low-Mg sequence in the [Mg/M] vs. [M/H] space, and few stars are located in the high-Mg population. More specifically, the ratio between the number of low-Mg and high-Mg stars increases when considering the outer Galactic regions.

\section{The GCE model for the disk and \texttt{OMEGA+}}\label{omega}

In this Section, we present the main properties of  our GCE simulation presented here, confirming and expanding the main results by \citet{spitoni_2019, spitoni_2021}. We describe how the galaxy evolution code works, along with the modifications to the original \texttt{OMEGA/OMEGA+} software.

\begin{table*}
\caption{List of the GCE parameters.}
\label{tab:help}    
\centering
\begin{tabular}{ccc}
\hline        Symbol & Definition & Fitting value or range\\
\hline\hline  $\nu$ & star formation efficiency, SFE &  0.022\\
        $\eta$ & mass-loading factor & 1 \\
         $t_{\beta}$ [Gyr] & characteristic decay time in the delay-time distribution of SNe~Ia & 2\\
        $N_{\rm Ia}/M_\odot$  &  number of SNe~Ia per stellar mass formed in a simple stellar population & $0.0020$\\
        $M_{\rm trans}$ [$M_{\odot}$] &  initial mass, which marks the transition from AGB to massive stars & 9\\
        $t_{\rm max, 1}$ [Gyr] & time of the first infall event & 0.10\\
        $t_{\rm max, 2} = t_{\rm max}$ [Gyr] & time of the second infall event & [1,12]\\
        $\tau_1$ [Gyr] & characteristic time of the accretion after the first infall event & [0,7]\\
        $\tau_2$ [Gyr] & characteristic time of the accretion after the second infall event & [0,10]\\
        $\tau'_{1}$ [Gyr] & characteristic rising time of accretion before the first infall event & 0.05 \\
        $\tau'_{2} = \tau_{\rm up}$ [Gyr] & characteristic rising time of accretion before the second infall event & [0,5] \\
        $\sigma_1$ [$M_{\odot}$pc$^{-2}$] & surface mass density of the gas accreted during the first infall event & $\sigma_2/\sigma_1 = [0.1,50]$\\
        $\sigma_2$ [$M_{\odot}$pc$^{-2}$] & surface mass density of the gas accreted by the merger event & $\sigma_1 + \sigma_2 =$[20,500]\\
\hline
\end{tabular}
\tablefoot{The third column contains the fixed value or the fitting range of each parameter.}
\end{table*}

\begin{table}
\caption{The present-time values of the global, observable parameters.}
    \centering
    \begin{tabular}{cccc}
\hline        Global observable && Value & Ref. \\
\hline\hline  SFR [M$_{\odot}$yr$^{-1}$] && 1.54 $\pm$ 0.56 & RW10\\
        $\dot{M}_{\rm in}$ [M$_{\odot}$yr$^{-1}$] && 1.1 $\pm$ 0.5 & M+12, LH11\\
        $M_{\star}$ [$10^{10}~{\rm M_{\odot}}$] && 3.5 $\pm$ 0.5 & F+06 \\
        $M_{\rm gas}$ [$10^{10}~{\rm M_{\odot}}$] && 0.81 $\pm$ 0.45 & K+15 \\
        rate of SN~Ia [$100^{-1}$ yr] && 0.4 $\pm$ 0.2 & P+11 \\
        rate of SN~II [$100^{-1}$ yr] && 2 $\pm$ 1 & P+11 \\
\hline
\end{tabular}
\tablebib{
RW10: \citet{RW10}; M+12: \citet{M12}; LH11: \citet{LH11}; F+06: \citet{F06}; K+15: \citet{K15}; P11: \citet{P11}.
}
\label{tab:present_values}
\end{table}

\subsection{Fundamental assumptions of Galactic Chemical Evolution}\label{assum}

\subsubsection{Distribution of individual stars} \label{yield_def}

To describe the $dN$ number of stars in the interval $[m, m + dm]$, we adopt \citet{kroupa_2001} for the initial mass function:
\begin{equation}\label{eq:imf}
    dN = \xi_0 m^{-a(m)} dm
\end{equation}
which has a power-law index of $a_1 = 1.3 $ over the range of stellar masses $0.08 \leq m/M_{\odot} < 0.5$, and $a_2 = 2.3$ otherwise. The $\xi_0$ normalization constant is obtained from the condition of $\int_{M_{\rm min}}^{M_{\rm max}}m\xi(m) dm=1$. 
Our prior tests suggested that the characteristic shape of the model curve in the [M/H]$-$[Mg/M] parameter plane was obtained with these choices of $a_1$ and $a_2$. We set the lower and upper mass limits of the IMF to 0.08~M$_{\odot}$ and 130~M$_{\odot}$, respectively.

The age of each source is crucial to define the timeline for mixing the metals in the ISM. On the other hand, massive stars effectively release metals instantaneously, when the simulation timestep is larger than their lifetimes. The number of explosions per stellar mass formed can be easily calculated from the IMF within the pre-defined mass range for massive stars of 9~M$_{\odot}$ to 30~M$_{\odot}$. Above the upper mass limit, we assume stars directly collapse into black holes \citep{ebinger_2020, boccioli_2024}. We acknowledge that this assumption may be incorrect in detail, as neutrino driven models of core-collapse supernovae (CCSNe, SNe type II) produce complex mass landscapes of black hole formation and successful explosions \citep[e.g., ][]{ugliano_2012,pejcha_2015}.
Conversely, progenitors of SNe~Ia have longer lifetimes within the \texttt{OMEGA+} simulations. As proposed by \citet{greggio_2005}, the occurrence rate of SNe~Ia for a simple stellar population (stars born at the same time and with the same chemical composition, SSP) can be calculated as the product of the SFR and a so-called delay-time distribution (DTD) function that describes the probability distribution of the explosion times. In this function, the fraction of white dwarfs is calculated from the lifetime of intermediate-mass stars and the IMF. The DTD($t$) function is normalized over the stellar lifetimes to describe the sequence of SNIa explosions. Therefore in the simulations, we applied the exponential approach introduced by \citet{pritchet_2008} and discussed by \citet{wiersma_2009}, where the decreasing DTD($t$) is written in the form of
\begin{equation*}
    {\rm DTD} ~\propto~ e^{-t/t_{\beta}},
\end{equation*}
where $t_{\beta}$ determines the characteristic decay time of the DTD. We set $t_{\beta} = 2\cdot 10^9$~years, as used by e.g., \citet{wiersma_2009}. 
The number of SNe Ia per stellar mass formed in a simple stellar population is assumed to be $N_{\rm Ia}/{\rm M}_\odot = 0.0020$ in our work, 
which was also proposed by \citet{maoz_2012} based on their examination of SN rate compilations.

\begin{figure}
\centering
\includegraphics[width=.5\textwidth]{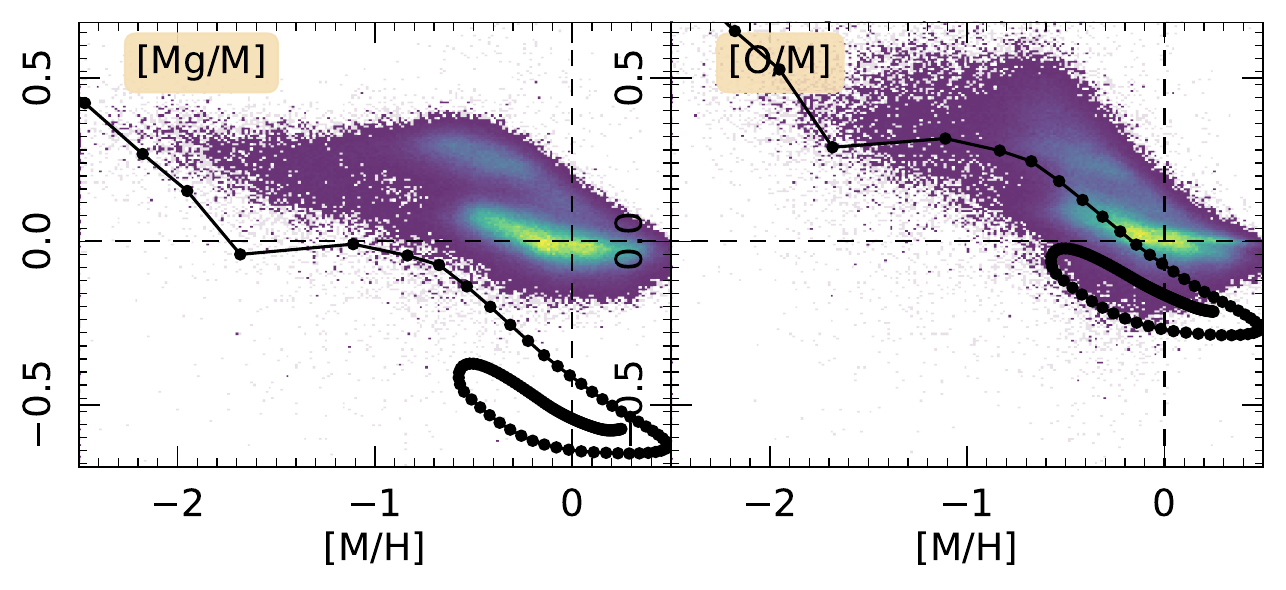}
\caption{A version of the GCE results for [M/H] versus Mg and O abundance ratios obtained without applying yield correction factors. The other GCE parameters are fixed at the best-fit values found in Sect.~\ref{res}.}
\label{fig:noyield}
\end{figure}

\subsubsection{The chemical evolution model for the Milky Way}\label{om_intro}

The time-delay model explains the behavior of the combined $\alpha$-elemental (O, Mg, S, Si, Ca) abundances because, in the early phases of Galactic evolution, these elements are only produced by short living massive stars, thus resulting in an average overabundance at low metallicity ($\mh<-1$~dex). Later SNe type Ia started to contribute with a time delay relative to CCSNe, and produced the bulk of the iron. As SNe type Ia begin to contribute, the [$\alpha$/M] ratios starts to decrease with time, until they reached the solar value. The main equation of chemical evolution describes the surface mass density of each element in the interstellar medium (ISM) at a given time. This may be solved only numerically, as the term involving the SFR and IMF grows with the integration time \citep[see e.g.,][]{matteucci_1985, matteucci_1986, francois_1986}. 

For the semi-analytical calculations, we use a modified version of \texttt{OMEGA+} \citep{cote_2018}, that is built on \texttt{OMEGA} \citep{cote_2017}. In the framework of \texttt{OMEGA+}, the simulated system consists of a cold gas reservoir, which is a star-forming region, and contains the stellar population of the Galaxy. This volume is surrounded by a hot gas reservoir (as the outer zone) filling the dark matter halo. This is considered to be the circumgalactic medium (CGM) from which the inflows happen. 
\texttt{OMEGA} simulates the inner star-forming region, then the rates of galactic inflow and outflow, and star formation are controlled by \texttt{OMEGA+}. 

The accretion from the external zone into the CGM is generally called the circumgalactic inflow and is assumed to have a primordial chemical composition. At the beginning of the simulation, the Galaxy is considered primordial. 
Then, the overall gas circulation, like the evolution of the mass of the cold gas reservoir, is tracked as the core of the calculations. The metallicity of the galactic gas is diluted by inflows from the CGM, stellar ejecta contributes to the mass-loss rate, and the star formation rate drives how much metal is ejected into the CGM. In each timestep, a simple stellar population is created that drives the galactic outflow, also resulting in a decrease of the inner metallicity \citep{cote_2018}.

In the default setup \citep{cote_2018}, the rate of the infalling matter to the galaxy from CGM is constant. We discuss in Sect.~\ref{om_impl} that the two-infall scenario instead requires an exponentially driven two-peak inflow rate, therefore we customized this function with free parameters while fitting.

\subsection{Implementations to \texttt{OMEGA+}}\label{om_impl}

To describe the relationship between the SFR and surface mass density of the gas, we adopt the Schmidt-Kennicutt law:
\begin{equation}\label{eq:kennicutt}
    \dot{M}_\star (t) = \nu \sigma^k (t),
\end{equation}
where $\nu$ is the star formation efficiency (SFE), which shows the fraction of the gas mass turned into newly forming stars per unit time, and $k=1.5$ \citep{schmidt_1959,larson_1988,larson_1992,kennicutt_1998}. We set $\nu = 0.022$ for the simulations.

Within the classical double-infall approaches \citepalias[e.g., ][]{spitoni_2021}, the inflow rate describing the merger episode is assumed to be zero until a certain time ($t_{\mathrm{max}}$) when the inflow peak is introduced instantaneously by an exponentially decreasing function with a characteristic accretion time $\tau$. 
To fine-tune our two-infall scenario, we introduce rising phases of the surface mass density arriving at the Galactic disk per unit time ($\dot{\sigma}(t)$): 
\begin{equation}
\begin{split}
        \dot{\sigma} (t) :=& \dfrac{\dot{M}_{\rm in}(t)}{A} = \sum_{k = \{ 1,2\} } \dot{\sigma}_{k,0} \bigg[ \theta (t_{\mathrm{max}, k} - t) \, e^{(t - t_{\mathrm{max}, k})/{\tau'_k}} \\ &+
    \theta (t - t_{\mathrm{max}, k}) \, e^{-(t - t_{\mathrm{max}, k})/{\tau_k}} \bigg]~,
\end{split}
\label{eq:sigmadot}
\end{equation}
where $A = (R^2 - r^2)\pi$ is the surface of the Galactic annulus bounded by $r$ and $R$, the $\dot{\sigma}_{k,0}$ normalization factors are the rates of the surface mass densities related to the $k$-th infall event, and $\theta(t-t_{\mathrm{max}})$ is the unit step function which has a value of zero if $t<t_{\mathrm{max}}$ and one otherwise. We implemented this form of inflow rate in \texttt{OMEGA+}. It provides infalling peaks that have an exponentially increasing phase and a decreasing branch with characteristic times $\tau_k'$ and $\tau_k$, respectively, both assumed as free parameters. More specifically, $\tau_1'$ corresponds to an intense infall at the beginning of galaxy formation. It has a fixed value of $\tau_1' = 0.05$~Gyr, while the other accretion parameters ($\tau_2'\equiv \tau_{\rm up}$, $\tau_1$, and $\tau_2$) are varied within the simulations.  
The time of the first infall is at the birth of the Galaxy, and for numerical reasons, it is fixed to be $t_{\rm max, 1}=0.1$~Gyr, and the $t_{\rm max, 2}\equiv t_{\rm max}$ delay duration between the two infall episodes is also a free parameter.

The $\dot{\sigma}_{k, 0}$ normalization factors related to the first and second infalls in Eq.~(\ref{eq:sigmadot}) are derived by the following considerations. The present-day surface mass density is the integral of  Eq.~(\ref{eq:sigmadot}) over the entire time duration of the evolution from $t=0$ to $t=t_{\rm G}$:
\begin{equation}
    \begin{split}
       \sigma_k(t_{\rm G})  &= \int_0^{t_G} \hspace{-.2cm} \dot{\sigma}_k (t)   dt\\
       &= \dot{\sigma}_{k,0} \left( \int_0^{t_{\rm max, k}} \hspace{-.2cm} e^{(t - t_{\rm max, k})/\tau'_k} dt +\int_{t_{\rm max, k}}^{t_G} \hspace{-.2cm} e^{-(t-t_{\rm max, k})/\tau_k} dt\right)\\
       &= \dot{\sigma}_{k,0} \left(  \tau'_k - \tau'_k e^{-t_{\rm max, k}/\tau'_k} + \tau_k - \tau_k e^{(t_{\rm max, k} - t_{\rm G})/\tau_k}    \right),
    \end{split}
\end{equation}
where $t_{\rm G}$ is the entire galactic lifetime. Thus, the normalization factor for each infall episode is determined as
\begin{equation}
    \dot{\sigma}_{k, 0} = \dfrac{\sigma_k (t_G)}{\tau_k'\left( 1 - e^{-t_{\rm max, k}/\tau_k'} \right) + \tau_k \left( 1 - e^{(t_{\rm max, k} - t_G)/\tau_k} \right)},
    \label{eq:twoinf-normfactor}
\end{equation}
where $\sigma_k(t_{\rm G})\equiv \sigma_{k}$ is the present-day surface mass density related to the $k$th infall episode. Here the sum and the ratio of the surface densities $\sigma_{\rm tot}$ and $\sigma_2/\sigma_1$ are the free parameters to optimize, where $\sigma_1$ and $\sigma_2$ can be obtained by $\sigma_2 = \sigma_{\rm tot}/(1+\sigma_2/\sigma_1)$ and $\sigma_1 = \sigma_{\rm tot} - \sigma_2$.

Therefore, the inflow rate is defined by the product of the surface mass density infalling per unit time and the 2-dimensional surface of the projection of the annular disk in the ($R-\phi$) plane (or on the direction perpendicular to the direction of $Z$-axis indicating Galactic height), and according to Eq.~(\ref{eq:sigmadot}), one can rewrite that the global inflow rate is
\begin{equation}
\begin{split}
    \dot{M}_{\rm in} (t) = (R^2 - r^2) \pi &\sum_{k = \{1,2\}}  \dot{\sigma}_{k,0} \bigg[\theta (t_{\mathrm{max}, k} - t) \, e^{(t_{\mathrm{max}, k} - t)/{\tau_k'}} \\&+
    \theta (t - t_{\mathrm{max}, k}) \, e^{-(t - t_{\mathrm{max}, k})/{\tau_k}} \bigg],
    \label{eq:22-twoinf-inflow-rate}
\end{split}
\end{equation}
where $r=3 ~\kpc$ and $R=15 ~\kpc$ are the inner and outer radii of the Galactic disk.

The total present-day surface mass density at the Galactocentric distance $R$ can be described as
\begin{equation}\label{eq:sigma_distr}
    \sigma_{\rm tot} (R) = \sigma_{\rm tot, \odot} ~e^{-(R-R_{\odot})/R_d}, 
\end{equation}
where $\sigma_{\rm tot, \odot}$ is the total surface density observed in the solar neighborhood, and the scale-length of the disk is $R_d=3.5$~kpc \citep{spigiomat_2017}. 
We introduce the $\sigma_{\rm tot, n}$ partial surface mass densities that are assumed to be constant within each 2-kpc wide region. 
Assuming that the values of $\sigma_{\rm tot, n}$ are constant within each region but fulfill the criterium of Eq.~(\ref{eq:sigma_distr}), the ratio between the densities in the neighboring regions is $q = \sigma_{\rm tot, n+1} / \sigma_{\rm tot, n} = e^{-\Delta R/R_d}$, where $n=\{ 1;2;...;6 \}$. Supposing that the partial densities are weighted by the powers of this quotient $q$, as well as by the area of the relevant annular galactocentric ring, we obtain
\begin{equation}
    A_n = \pi\left[ (R_n+\Delta R/2)^2 - (R_n-\Delta R/2)^2\right]=2\pi R_n \Delta R,
\end{equation}
and the $n$th partial surface mass density in the form of a sequence
\begin{equation}\label{eq:seq}
    \sigma_{\rm tot, n} = \sigma_{\rm tot, 0}\left[ R_1 + (n-1)\Delta R \right] q^{n-1},
\end{equation}
where $\sigma_{\rm tot, 0}$ is the inverse normalization coefficient. As the summation of this sequence should return the total fitted value,  the $\sigma_{\rm tot, 0}$ factor can be derived by the relation
\begin{equation}
    \sigma_{\rm tot, 0} = \dfrac{\sigma_{\rm tot}}{\sum_{i=1}^{6}\bigg( \left[ R_1 + (i-1)\Delta R \right] q^{i-1}\bigg)}.
\end{equation}
Thereinafter, the $n$th partial surface mass density is calculated according to Eq.~(\ref{eq:seq}). In the solar neighborhood, the total baryonic mass surface density today is $\sigma_{\rm tot, \odot} = (47.1 \pm 3.4)~{\rm M_{\odot} pc^{-2}}$ \citep{mckee_2015}. Since outflows also happen, the total infalling surface mass density in the solar region ($7~\kpc\leq R < 9~\kpc$) is accepted to result in a different value in the simulations. Also, the surface density $\sigma_{\rm tot, 3}$ represents the average surface density within the Galactic ring centered at $8~\kpc$.

In summary, the total infalling matter during the simulation is spatially distributed into the regions, and we assume that within the $n$th region, the regional partial surface mass density, $\sigma_{\rm tot, n}$ is constant along its width of 2~kpc. The values at the centers of the regions ($R_n={4;6;...;14}~\kpc$) follow the exponentially decreasing distribution described by Eq.~(\ref{eq:sigma_distr}). Each calculated  $\sigma_{\rm tot, n}$ is listed in Table~\ref{sum_results}, and shown in Fig.~\ref{fig:res_sigmas}. Due to these implementations, the regions are coupled to each other, since the integral of the global inflow rate function over Galactic age equals to the sum of the integrals of the partial inflow functions. 

\begin{figure*}
\centering
\includegraphics[width=\textwidth]{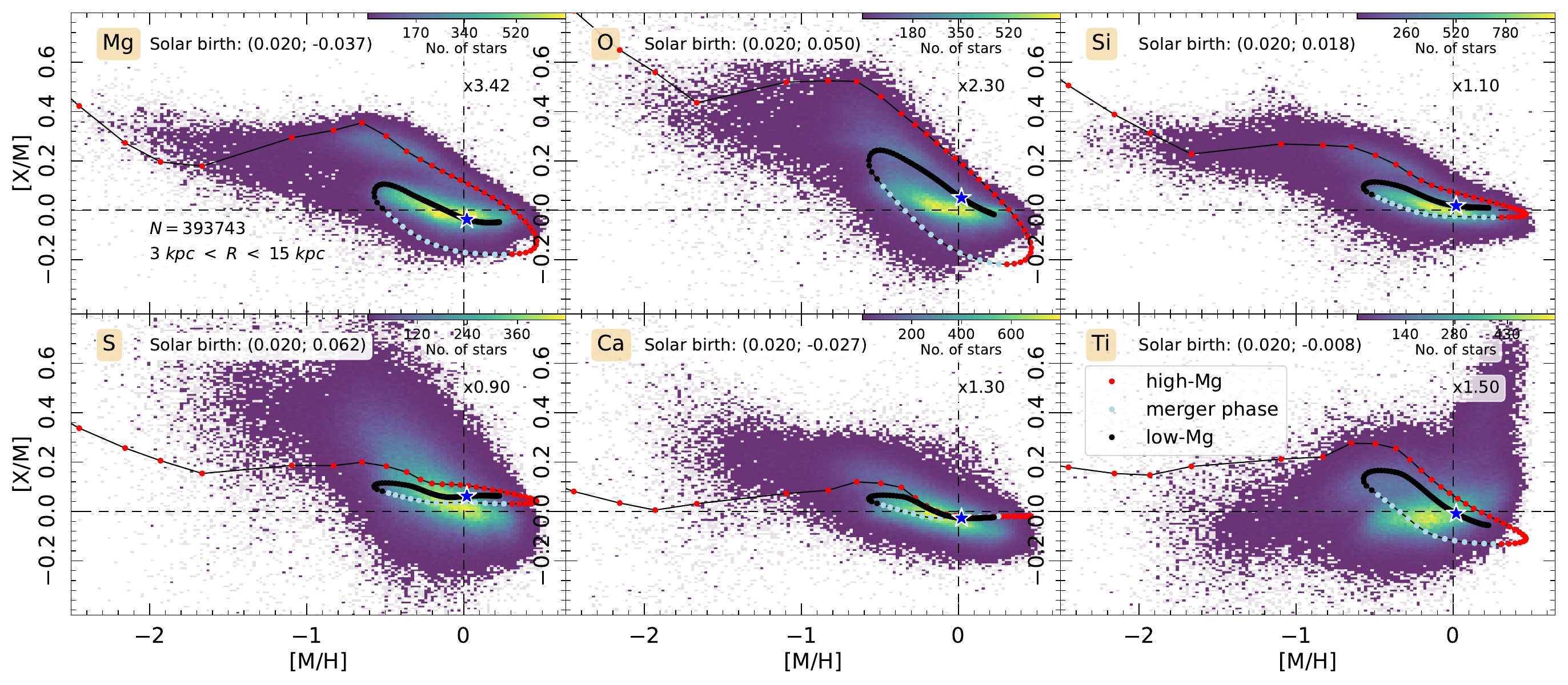}
\caption{Observed [X/M] vs. [M/H] abundance ratios for the $\alpha$-elements (O, Mg, Si, S, Ca, Ti) from MWM DR19 for the entire galactocentric region between 3 and 15~kpc compared with the best fit CE model results (dotted curves)  for all species. The spacing between the dots represent the discrete evaluation points during the time of the GCE simulation. The color coding represents the number of stars on the observational chemical map of the MW, and grey squares represent bins containing a single star. The red and black filled circles mark the path of the high-Mg and low-Mg stellar sequences, respectively, while the gray dots correspond to the so-called merger phase. This color coding is the same as in Fig.~\ref{global_fit_oddz} and the top panel of Fig.~\ref{present_v_global}. In each panel, the metallicity and [X/M] abundance ratio are displayed at the exact time of the Sun's birth during the simulation, which is also marked with a star symbol. The multiplication factor applied for the stellar nucleosynthetic yields of the relevant element is also indicated in the upper right corner of the plots.}
\label{global_fit}
\end{figure*}

At the beginning of the simulation and around the second infall event, GCE quantities (e.g., SFR) change rapidly. 
Therefore, the first timestep is $0.1~{\rm Myr}$ long, and until it reaches 50~{Myr}, a logarithmic scale is used. During the remaining Galactic time, a uniform resolution is applied. We consider the age of the Sun to be $4.57$~Gyr \citep[e.g., ][]{boothroyd_2003}.

\begin{figure}
\centering
\includegraphics[width=.5\textwidth]{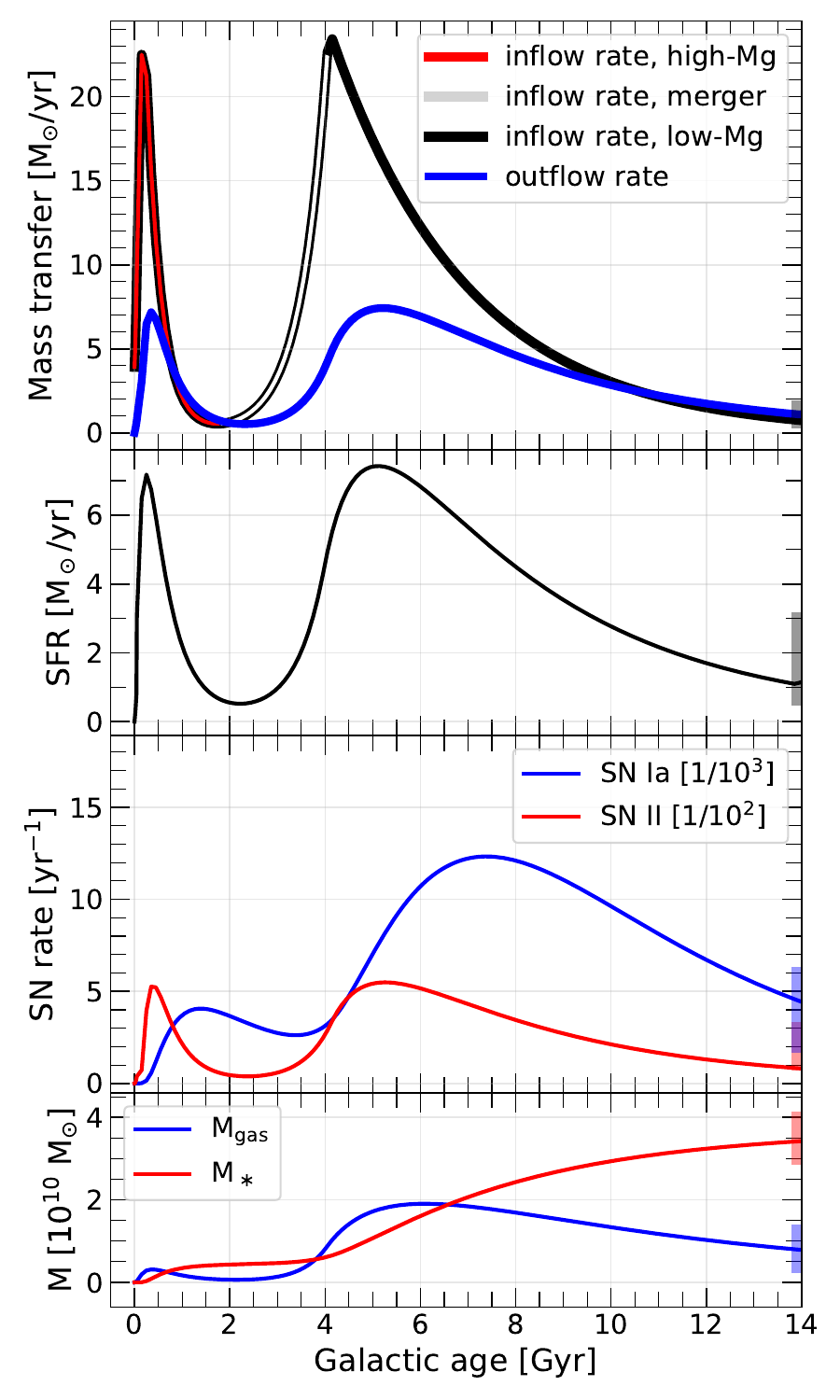}
\caption{The inflow and outflow rates (top panel), the star formation rate (second panel), the rates of SNe type Ia and II (third panel), the Galactic mass $M_{\star}$ enclosed in stars and the mass $M_{\rm gas}$ of the gas (bottom panel) as a function of galactic age, where $t=0$~Gyr is the beginning of the Galactic formation, and $t=14$~Gyr is present. The shaded areas with the corresponding colors represent the present-day measured values for each parameter (see Table~\ref{tab:present_values}). The inflow rate presented in the top panel includes the high-Mg (formation) and low-Mg (merger relaxation) phases, that are marked with red and black curves, while the rising accretion of the merger phase is marked with gray.}
\label{present_v_global}
\end{figure}

\begin{figure}
\centering
\includegraphics[width=.5\textwidth]{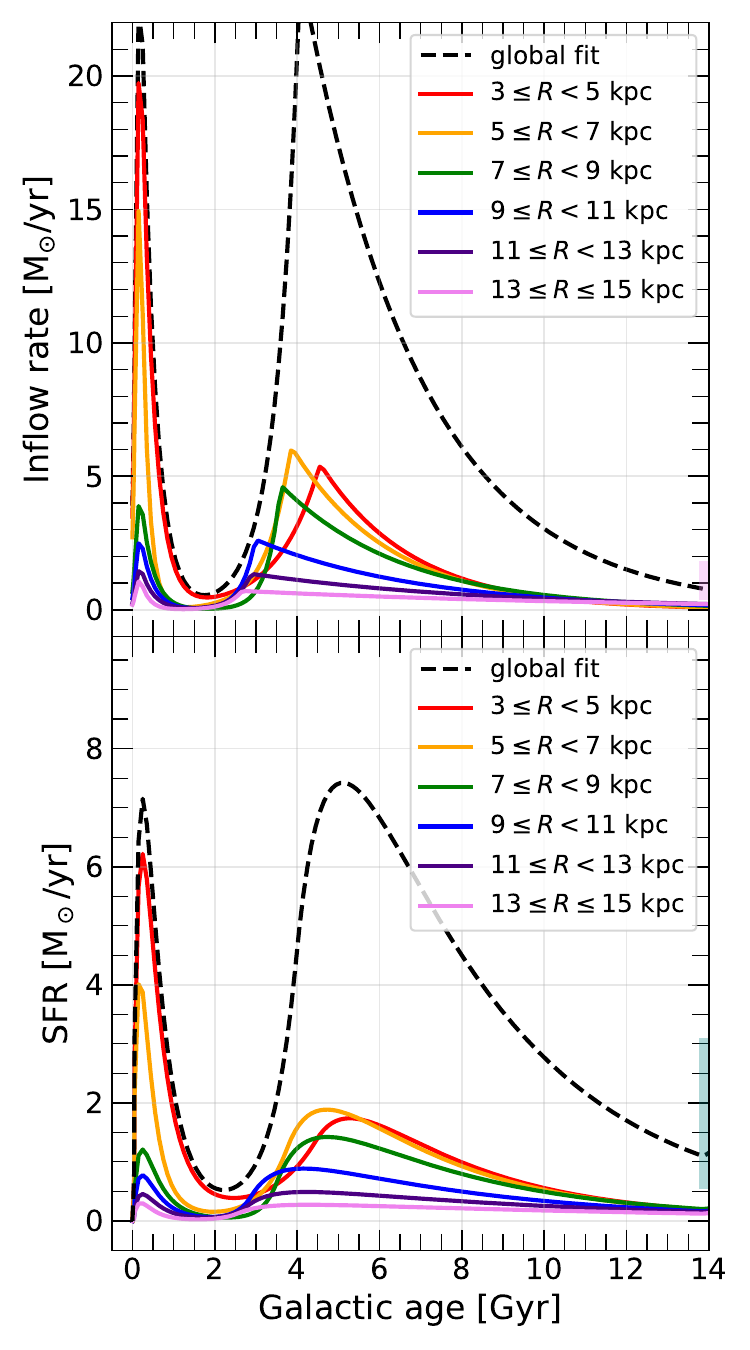}
\caption{\textbf{Top panel}. Inflow rates as a function of galactic age, where $t=0$~Gyr is the beginning of the Galactic formation, and $t=14$~Gyr is present. \textbf{Bottom panel}.The star formation rate as a function of time. The dashed and colored curves show the results of the global and the regional best-fit simulations, respectively.}
\label{fig:res_inflow}
\end{figure}

\subsection{Nucleosynthetic yield sets}
\label{sec:yields}

Depending on its initial mass, each star produces unique amounts of the chemical elements, and ejects them in the ISM once its lifetime expires. 
Stellar yields are computed with detailed nucleosynthesis calculations considering all the main nuclear reactions in stars and supernovae. Theoretical stellar yields are used as input for different GCE models. 
The mass of a newly formed element $k$ in a star of mass $m$ is defined as
\begin{equation}\label{eq:m_k}
    M_{km} = \int_0^{\tau_m}\dot{M}_{\rm lost}(t) \cdot \left[ X(t, k) - X_o(k) \right] dt,
\end{equation}
where $\tau_m$ is the lifetime of a star of initial mass $m$, $\dot{M}_{\rm lost}$ is the mass loss rate of the star (the rate of the mass that is ejected by the star into the ISM) and $X_o(k)$ and $X(t,k)$ are the original and final abundances of the element $k$, respectively. The stellar yield is given by the fraction of the stellar mass converted into that element: $p_{km} = M_{km}/m$, where $M_{km}$ is the total mass of the newly formed chemical species.

Through the JINA-NuGrid chemical evolution pipeline \citep{cote_2017b}, \texttt{OMEGA+} has access to NuPyCEE stellar yields library for low-, intermediate-, and high-mass stars \citep{pignatari_2016, ritter_2018}. The tables provide the stellar yields on a grid that spans over 12 stellar models 
between 1 and 25~M$_{\odot}$, and between five metallicities from $Z=0.0001$ to $0.02$. Beyond these points, stellar yields are interpolated as a function of metallicity and mass.

\begin{table*}
\caption{Results for the global and regional fits.}
\begin{tabular}{c|cc|cc|cc||c||c}
\hline\hline 
zone & \multicolumn{2}{|c|}{\textbf{inner}}  &  \multicolumn{2}{|c|}{\textbf{middle}}  &  \multicolumn{2}{|c||}{\textbf{outer}}  & this model & Ref. \citepalias{spitoni_2021}
\\
\hline
region  & 
{\textbf{R$_1$}} & {\textbf{R$_2$}} & {\textbf{R$_3$}} & {\textbf{R$_4$}} & {\textbf{R$_5$}} & {\textbf{R$_6$}} & \textbf{global} & \textbf{global}   \\
distance [kpc]& 
{$3-5$} & {$5-7$} & {$7-9$} & {$9-11$} & {$11-13$} & {$13-15$} & {$3-15$} & {$2-14$}\\ 
$N_{\star}$ & 20,255 & 48,802 & 171,134 & 100,054 & 39,531 & 13,967 & 393,743 & 26,690\\
\hline 

$t_{\rm max}$ [Gyr] 
& 4.55 $\pm$ 0.11 & 3.82 $\pm$ 0.10 & 3.65 $\pm$ 0.03 & 2.94 $\pm$ 0.03 & 2.84 $\pm$ 0.05 & 2.67 $\pm$ 0.16 & 4.13 $\pm$ 0.19 & 3.00$-$4.71 \\ 

$\tau_1$ [Gyr]
& 0.30 $\pm$ 0.02 & 0.16 $\pm$ 0.01 & 0.29 $\pm$ 0.01 & 0.29 $\pm$ 0.02 & 0.30 $\pm$ 0.02 & 0.20 $\pm$ 0.02 & 0.32 $\pm$ 0.02 & 0.10$-$0.47\\ 

$\tau_2$ [Gyr]
& 2.23 $\pm$ 0.13 &2.40 $\pm$ 0.12 &3.00 $\pm$ 0.15 &4.00 $\pm$ 0.28 &5.99 $\pm$ 1.19 &9.40 $\pm$ 2.35 & 2.86 $\pm$ 0.70 & 3.63$-$11.49 \\

$\sigma_2 / \sigma_1$
&1.99 $\pm$ 0.06 &4.34 $\pm$ 0.11 &8.80 $\pm$ 0.26 &9.86 $\pm$ 0.32 &11.49 $\pm$ 0.44 &14.29 $\pm$ 0.60 & 7.61 $\pm$ 0.23 & 3.69$-$10.54\\

$\sigma_{\rm tot}$ [$M_{\odot}$pc$^{-2}$] 
& 46.9 $\pm$ 8.8 & 39.8 $\pm$ 7.4 & 29.9 $\pm$ 5.6 & 21.1 $\pm$ 3.9 & 14.3 $\pm$ 2.7 & 9.4 $\pm$ 1.8 & 161.5  $\pm$ 30.2 & $-$ \\

$\tau_{\rm up}$ [Gyr]
& 1.00 $\pm$ 0.20 & 0.58 $\pm$ 0.07 & 0.32 $\pm$ 0.03 & 0.33 $\pm$ 0.03 & 0.40 $\pm$ 0.04 & 0.37 $\pm$ 0.07 & 0.55 $\pm$ 0.06 & $-$ \\
\hline 
\label{sum_results}
\end{tabular}
\tablefoot{The $\sigma_{\rm tot}$ total surface mass density is fitted in the global sample. The $\sigma_{\rm tot, n}$ partial surface mass densities are calculated by a weighted distribution within the six regions in a way described in the text. The number of stars contained by each zone is denoted by $N_{\star}$.}
\end{table*}

\begin{figure}
\centering
\includegraphics[width=.5\textwidth]{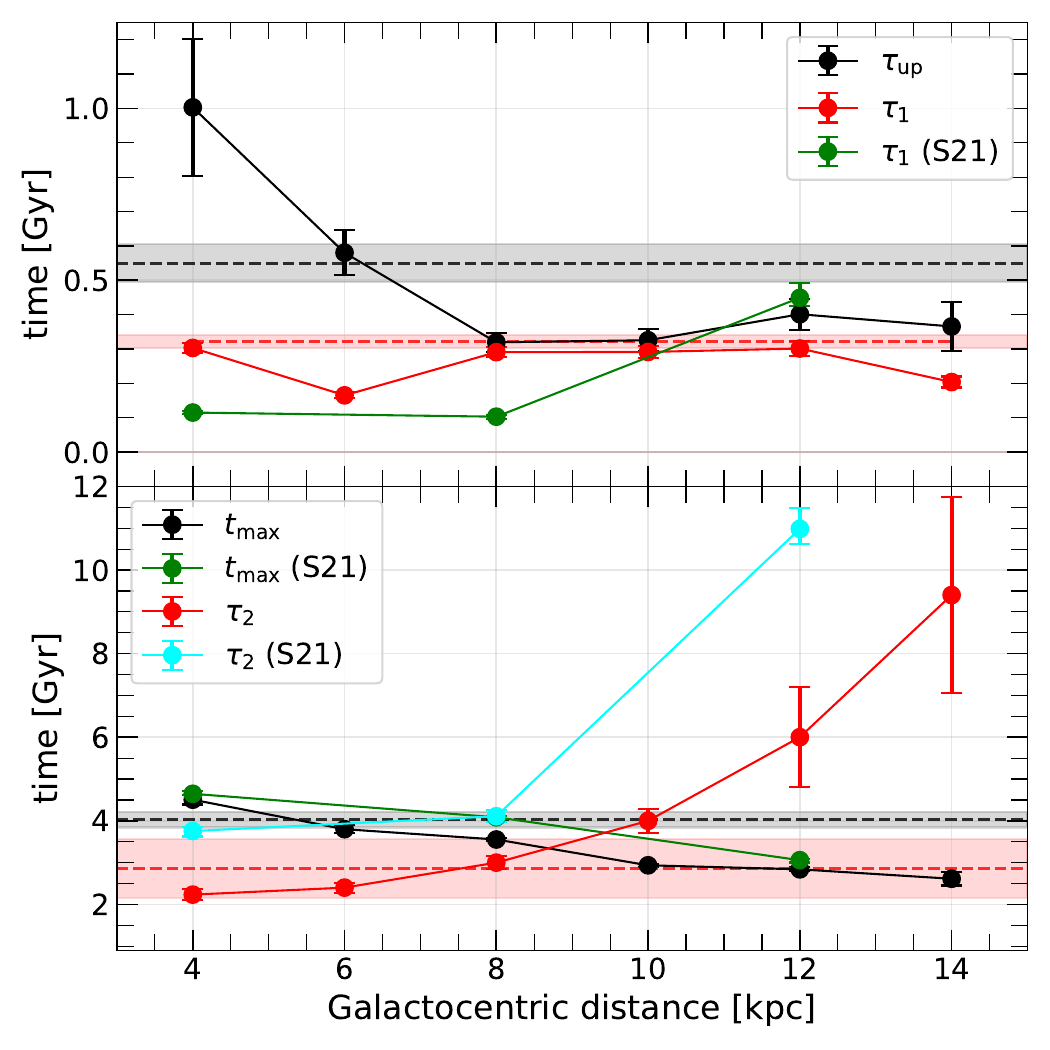}
\caption{Characteristic ascending and descending accretion timescales ($\tau_1, \tau_2, {\rm ~and~} \tau_{\rm up}$) and time delay ($t_{\rm max}$) from the best-fit results as a function of Galactocentric distance $R$ [kpc]. While the continuous lines represent the regional fits, the global value is indicated by dashed lines. {Error bars and shaded areas show the estimated uncertainties of the fitted global and regional parameters.} Results from \citetalias{spitoni_2021} are denoted with green and cyan. }
\label{fig:results_times}
\end{figure}

The initial metallicity of the gas in mass fraction of all stellar ejecta is uniformly set to be $Z=0.001$ in this work. We adopt the \citet{cristallo_2015} yield set within the regime up to the asymptotic giant branch. The AGB yields are extracted from FRUITY \citep{cristallo_2011} and massive star yields from NuGrid \citep{pignatari_2016}, whereas lifetimes for AGB models were taken from NuGrid \citep{ritter_2018}. The minimum mass of SNe type II ($M_{\rm trans}$) sets the transition between intermediate mass and massive stars. We set the value of $M_{\rm trans}=9~M_{\odot}$, consistent with the that used in literature \citep[$8-10~M_{\odot}$, ][]{matteucci_2021}. 
The yields of SNe Ia are taken from \citet{iwamoto_1999}. For the nucleosynthetic yields of an initial generation of massive metal-free stars (often referred to as population III stars), we adopt  \citet{heger_2010}. We assume that stars with initial masses of 10 to 100~M$_{\odot}$ contribute to explosive nucleosynthesis, while those more massive than 100~M$_{\odot}$ collapse to black holes.

The nucleosynthetic yields for many elements are under or over predicted due to many factors including uncertainties in nuclear reaction rates and atomic cross sections, explosion assumptions, and computational limitations. To correct for these known problems, we introduce multiplication factors, denoted by $f_X$, to the values reported in the yield tables. These correction factors are fit as free parameters in our final GCE models. To demonstrate the necessity of the correction factors, in Fig.~\ref{fig:noyield} we show the global GCE model results for Mg and O without the $f_X$ corrections compared to the MWM data. We see that the model severely under-predicts [Mg/M] and slightly under-predicts [O/M] when no yield correction factors are applied. We fit one correction factor per element, and multiply yields from all nucleosynthetic sources by this value.

\section{Fitting methods}\label{sec:methods}

\subsection{Determination of the free parameters in the GCE models}\label{sec:free-and-fixed}

The main GCE parameters are listed and defined briefly in Table~\ref{tab:help}. We categorize the model parameters and the effects of physical processes into four groups based on how we treat them in the fitting method. First, there are effects that we ignore, such as dark matter in the disk, pre-enriched inflows, variation of the SFE and mass-loading factor ($\eta$) along the Galactocentric radii, the lifetime of stars above $M_{\rm trans}$, and the effect of radial stellar migration. 

Secondly for the GCE simulations, we fix the $M_{\rm trans}=9~M_{\odot}$ lower boundary of massive stars, and the [0.08, 130] range of initial mass function in solar masses. Consistently with the considerations of \citet{cote_2016}, we assume that above the mass of $M_{\rm th} = 30~M_{\odot}$ stars do not release any ejecta, but directly collapse into black holes. Therefore, the maximum mass in the IMF affects the total mass of gas locked into stars, but not the ejected yields. In order to test such an assumption, we have made additional GCE simulations using $M_{\rm th} = 100~M_{\odot}$ as an upper mass limit, and we obtain no significant changes in our results presented here. In particular, the increase of the upper mass limit  causes a variation mainly in the set of $f_X$ values for the $\alpha$-elements (e.g., 12\% decrease for Mg), while the relevant parameters of our MW model presented in Table~\ref{sum_results}. vary by less than $10\%$: $4-8\%$ for $\sigma_2 / \sigma_1$, $\tau_{\rm up}$ and $\tau_1$, and around $8-9\%$ for $t_{\rm max}$ and $\tau_2$. There are no big variations in the other relevant MW parameters of the model.

Third, we performed initial tests to set the following parameters and not involve them later in the final fitting: the $\eta$ mass-loading parameter, the $t_{\beta}$ exponent in the DTD function, the star formation efficiency parameter $\nu$ (see later in Eq.~(\ref{eq:kennicutt})), the $N_{\rm Ia}/M_\odot$ number of SNe Ia per stellar mass formed in a simple stellar population, the $\tau'_{1}$ first infall characteristic time of the rising branch, and the $a$ slope of the IMF. These prior tests provided the distinct shape of the model curve. Moreover, another aim of the initial \texttt{OMEGA+} tests was to ensure that the evolutionary curves (see later in, e.g., Fig.~\ref{global_fit}) for the $\alpha$-elements (in Fig.~\ref{global_fit_oddz}) and for the odd-Z elements intersect the zeros on both axes [M/H] and [X/M] at the time of the birth of the Sun. 

\begin{figure}
\centering
\includegraphics[width=.5\textwidth]{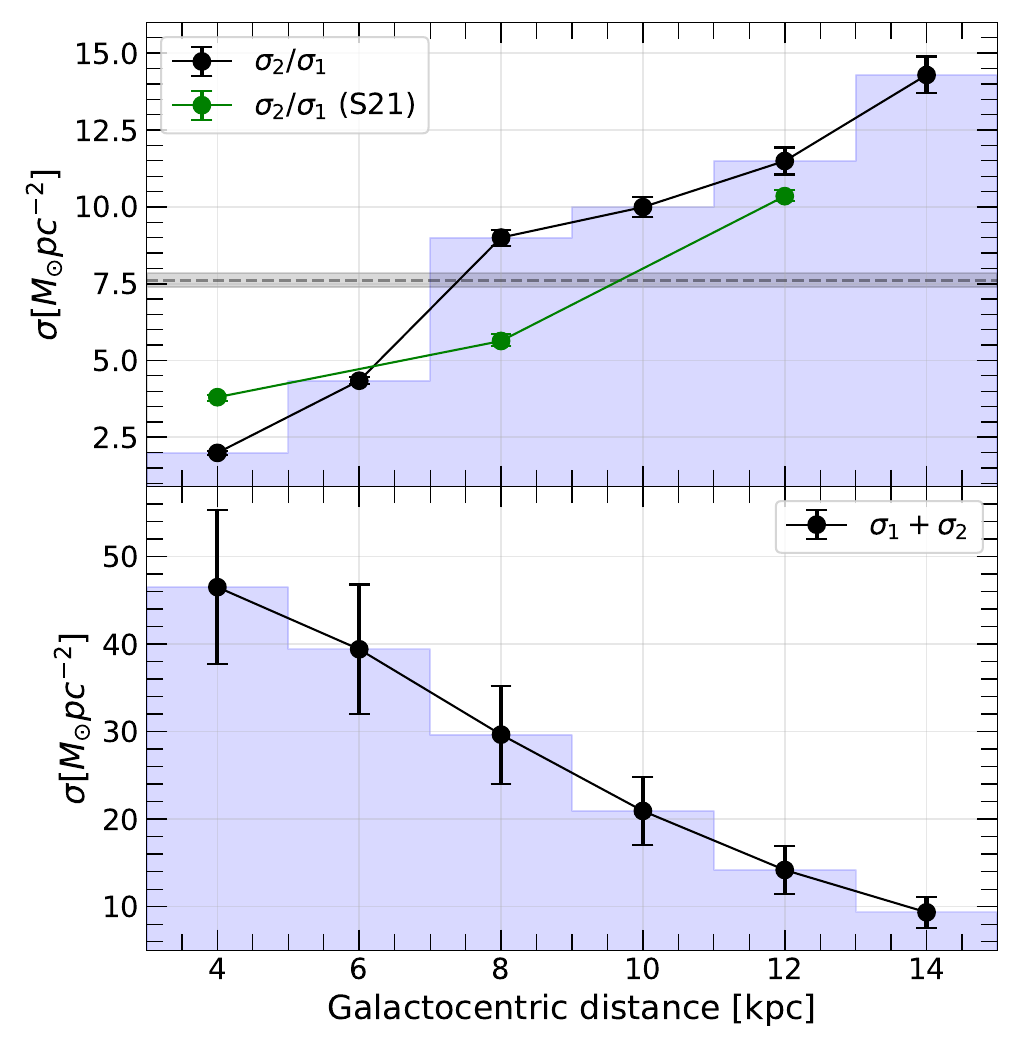}
\caption{The ratio (top panel) and the sum of the surface mass densities (bottom panel) related to the second and first infall events as a function of Galactocentric distance $R$ [kpc]. The dashed line in the top panel represents the global fitted value, and the shaded area is the related uncertainty. Note that the errors associated with the total regional surface mass densities are calculated as described in the text. Results from \citetalias{spitoni_2021} are denoted with green. }
\label{fig:res_sigmas}
\end{figure}

\begin{figure}
\centering
\includegraphics[width=.5\textwidth]{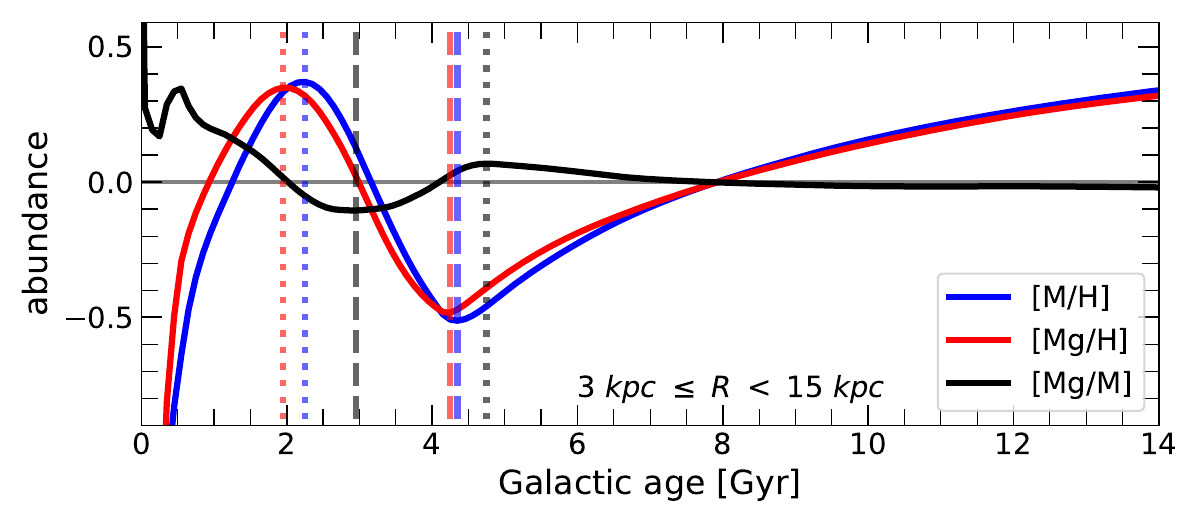}
\caption{Evolution of the [M/H], [Mg/H], and [Mg/M] abundance ratios as a function of Galactic age [Gyr] are marked with blue, red, and black colors, respectively. Note that a conversion relation of [Mg/H] = [Mg/M] + [M/H] holds for the values. The curves represent the result of the best-fit global model. Vertical dotted and dashed lines respectively represent the maxima and minima of each abundance curve within the range of $1-7$~Gyr, where $t=0$~Gyr is the beginning of the Galactic formation, and $t=14$~Gyr is present.}
\label{fig:feh_vs_age}
\end{figure}

The normalization of the GCE curves to the solar composition is not ideal. However, in our analysis we will look more specifically at the abundance patterns with respect to metallicity, which are not modified by forcing the curves to fit the Sun. It is clear, however, that successful GCE simulations of the solar neighborhood should be able to reproduce the solar abundances \citep[e.g., ][]{pignatari_2023}, according to Eq.~(\ref{eq:abund_def}). In addition, these tests tightened the fitting ranges of the free parameters. They showed what setups we should not consider further due to not meeting the present values of observable parameters of the MW within $\pm$ 100\%. These present-time values of the global, observable parameters are listed in Table~\ref{tab:present_values}.

\begin{figure}
\centering
\includegraphics[width=.485\textwidth]{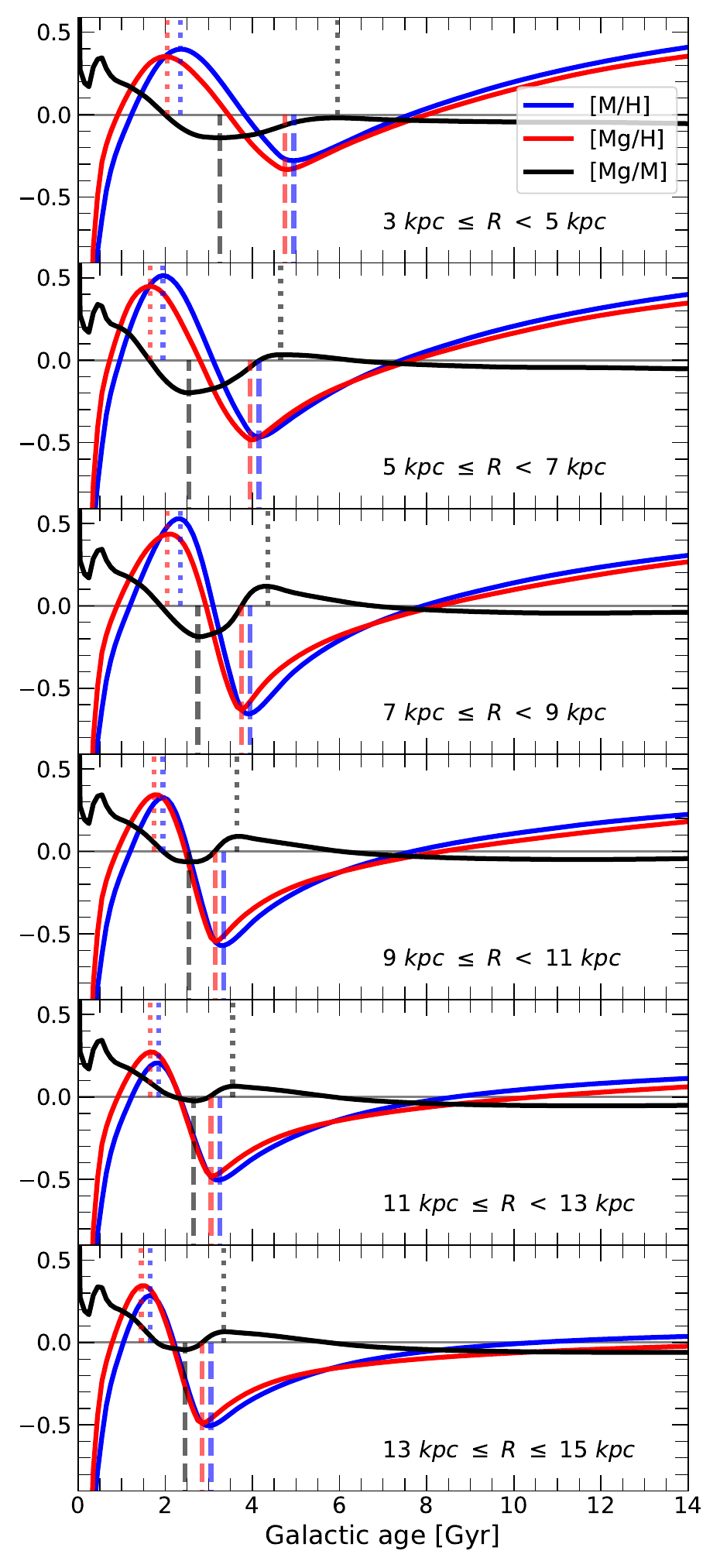}
\caption{Same as Fig.~\ref{fig:feh_vs_age}, representing the regional cases computed at $R=\{4, 6, 8, ..., 14\}$~kpc.}
\label{fig:feh_vs_age_regional}
\end{figure}

Finally, we fit the following free parameters in our models: $t_{\rm max}$, $\tau_{\rm up}$, $\tau_1$, $\tau_2$, $\sigma_{tot}$, $\sigma_2/\sigma_1$,  and $f_{X}$. The time elapsed between the birth of the Galaxy and the second infall is denoted by $t_{\rm max}$, whose prior range was $1 \leq t_{\rm max} \leq 12$~Gyr. The $\tau_{\rm up}$ is the characteristic time of the ascending branch connected to the second infall event, and we set it to span between $0 \leq \tau_{\rm up} \leq 5$~Gyr. The accretion timescales associated with the first and second infall events, $\tau_1$ and $\tau_2$, have varying ranges of $0 \leq \tau_{1} \leq 7$~Gyr and $0 \leq \tau_{2} \leq 10$~Gyr, respectively. The sum of the surface mass densities resulted from the first and second infall events is $\sigma_{tot}=\sigma_{1}+\sigma_{2}$. We set a prior fitting range of $20 \leq \sigma_{\rm tot} \leq 500$~M$_{\odot}$pc$^{-2}$. In contrast, we allow the ratio of these two surface densities to vary within values $0.1 \leq \sigma_2/\sigma_1 \leq 50$. According to our primary simulations, modifying the values in the isotopic yield tables was necessary, as using the original values do not fit the observations (see in Fig. \ref{fig:noyield}). 
All the parameters mentioned here are listed in Table~\ref{tab:help}, along with their fixed values or varying intervals.

\subsection{Fitting the data with \texttt{lmfit} algorithm}\label{fitting}

To choose the best-fit model, we create a pipeline that starts with the separation of the observational data into the high- and low-Mg populations and finishes with reporting the optimized GCE model via the nested \texttt{OMEGA+} code. The \texttt{lmfit} Python package contains the fitting method solving for least squares with the Levenberg-Marquardt (LM) algorithm. This package provides tools to build complex fitting models for non-linear least-squares problems and to apply these models to the measured data. 
The $\chi^2$-statistic is defined as:
\begin{equation}\label{chi}
    \chi^2 = \sum_i^{N}\dfrac{\left[ y_i^{\rm meas} - y_i^{\rm model} (\textbf{p})\right]^2}{\epsilon_i^2} ~,
\end{equation}
where $N$ is the number of values at which the function is evaluated, 
$y_i^{\rm meas}$ is the set of measured data, 
$y_i^{\rm model}(\textbf{p})$ is the model, 
$\textbf{p}$ is the set of variables in the model to be optimized in the fit, and $\epsilon_i$ is the estimated uncertainty in the data. 
This method is based on an objective function that takes a set of variables, then calculates the model and returns a residual array of $y_i^{\rm meas} - y_i^{\rm model}(\textbf{p})$. To perform the parameter optimization, a minimization is carried out on the residuals with \texttt{scipy.optimize}\footnote{\url{docs.scipy.org/doc/scipy/reference/optimize.html}}. After a fit has been completed successfully, the estimated uncertainties for the fitted variables and correlations between pairs of fitted variables are calculated by inverting the Hessian matrix, which represents the second derivative of fit quality for each variable parameter. 

The method of dividing the locus of high- and low-Mg populations from the observed abundances is explained in Sect.~\ref{mod_descr}. In each iteration phase, the observed median high-Mg and low-Mg sequences are interpolated over the output \mh ~values. This intrinsically gives higher weights to the data points that belong to the thin disk, as the metallicity steps during the GCE simulation get smaller when more stars are created within a timestep. 
The [M/H] vs. [Mg/M] parameter spaces then correspond to the 
\begin{equation*}
    \textbf{w}_{\rm global} = \{ w_{\rm \mh, global}, w_{{\rm [X/M], global}} \}  ~, ~~~{\rm and}
\end{equation*}
\begin{equation*}
    \textbf{w}_{n} = \{ w_{\mh,n}, w_{{\rm [X/M]}, n} \}
\end{equation*}
arrays featuring the global and the regional data (describing the $n$th region, where $n\in\{ 1;2;...;6 \}$), respectively. The $w_{\rm [X/M], global}$ and $w_{{\rm [X/M]}, n}$ notations provide the $y^{\rm meas}$ values within Eq.~(\ref{chi}). 
Within our analysis, the data$-$model residual returned by the objective function is the difference between the observational data and the model, where the latter is based on the output of the GCE code. The results of an \texttt{OMEGA+} run contain all the relevant quantities to analyze the GCE model (e.g., Galactic gas mass, SNe type Ia and II rate, elemental abundances).

The global and regional parameters that are varied are
\begin{equation*}
    \textbf{P}_{\rm global} = \{ t_{\rm max}, \tau_1, \tau_2, \tau_{\rm up}, \sigma_2/\sigma_1, \sigma_{\rm tot}, {f_{\rm Mg}} \} ~, ~~~{\rm and}
\end{equation*}
\begin{equation*}
    \textbf{P}_{\rm regional} = \{ t_{\rm max}, \tau_1, \tau_2, \tau_{\rm up}, \sigma_2/\sigma_1\} ~,
\end{equation*} 
where each parameter spreads within a range detailed in Table~\ref{tab:help}. These vectors are considered the set of variables $\textbf{p}$ in Eq.~(\ref{chi}).

This study is computationally built upon three main sequential blocks. Initially, we perform a global fit on the galactic sample by optimizing the $\textbf{P}_{\rm global}$ parameters to fit the [Mg/M]-[M/H] parameter plane. We use Mg as the key tracer as it is one of the most accurately measured and precisely derived elements published in MWM, and its zero-point offset is negligible for RGB stars. The estimated MWM precision varies between 0.04 and 0.05~dex (M25). Then, in the second phase, the optimal global data set is fixed, but we allow the yield multiplication factors, $f_{\rm X}$, to vary for the other chemical species to match the [X/M] observations. Finally, the regional fits are optimized by fixing the global parameter $\sigma_{\rm tot}$ as well as $f_X$ while fitting the $\textbf{P}_{\rm regional}$ variables. 

Note that the distributions of the components of the residual array are approximated as Gaussians, because investigating the correlations and degeneracies between the GCE parameters would lead us beyond the scope of this study. The errors of the $\textbf{P}_{\rm global}$ and $\textbf{P}_{\rm regional}$ parameters are reported by the LM-algorithm. However, the uncertainties of the resulting present values of the inflow rate, SFR, SNe rates, and the gas and stellar mass of the Galaxy must be estimated with a different approach, as they are not free parameters. Therefore, we implement a boot-strap analysis, which involves 10,000 random re-samplings of the observational data, and the distributions of the obtained values are approximated with Gaussians. Then the $1\sigma$ standard variations provide estimates of the uncertainties.  

The best-fit GCE model parameters with confidence levels are reported in Table~\ref{sum_results} both in the global and regional cases. The yield multiplication factors and the discrepancies of $\Delta \mh$ ~and $\Delta$[X/M] from the solar zero-point at the time of the birth of the Sun are summarized later in Table~\ref{tab:sum_yields}. The solar reference abundance scale used in MWM is \citet{grevesse_2007}, therefore, we implemented the same table in the GCE model to normalize the chemical abundances with.

\section{Results and discussion} \label{res}

The model GCE predictions are obtained by fitting the abundance ratios [Mg/M] and [M/H] of stars in the MWM DR19 data set using the \texttt{lmfit} technique. The global fits are discussed in Sect.~\ref{global}, and in Sect.~\ref{sec:along_radius} we interpret our results in terms of the inside-out formation scenario. In Sect.~\ref{comparison}, we compare our findings with the predictions of \citetalias{spitoni_2021}.

The first panel of Fig.~\ref{global_fit} shows the observed [Mg/M] vs. [M/H] abundance ratios derived in MWM DR19 within the disk between 3 and 15~kpc compared with the best-fit GCE model results (dotted curves). Note that this parameter plane was used for obtaining the best-fit global GCE model. Figures \ref{present_v_global} depicts the inflow and outflow rate, SFR, SNe rates, and the evolution of gas and stellar mass according to the global model, while Fig.~\ref{fig:res_inflow} is linked to the regional results. The characteristic ascending and descending accretion timescales and the galactic times of the second infall are plotted in Fig.~\ref{fig:results_times}, the ratios and the sums of the surface mass densities are represented in Fig.~\ref{fig:res_sigmas}. The evolution of the individual $\alpha$-elements, odd-Z elements, and metallicity are tracked in Figs.~\ref{global_fit},\ref{global_fit_oddz}, and \ref{fig:feh_vs_age}-\ref{fig:feh_vs_age_regional}, respectively.

\subsection{Global evolution of the Galactic disk}\label{global}

The evolution of the global observables (inflow and outflow, SFR, SNe rate, galactic and stellar mass) is given in Fig.~\ref{present_v_global}. 
The final quantities obtained at the end of the simulations are called present-day values, and their final values are compared to the literature values derived mostly based on solar neighborhood observations and listed in Table~\ref{tab:present_values}.

The inflow and the delayed outflow trends are plotted in the top panel of Fig.~\ref{present_v_global}. The maximum inflow rate during the formation phase, which happened in the first 2 Gyr, is $28.72$~M$_{\odot}$yr$^{-1}$ at $0.15$~Gyr, while the maximum outflow rate during this period is $8.58$~M$_{\odot}$yr$^{-1}$ at $0.35$~Gyr. This high-Mg phase is followed by the ascending (accretion) branch of the merger event, in which the inflow rate sharply increases with time after a short gap. In the gap, the inflow and outflow rates reached their minimum at around $1.45$~Gyr and $2.05$~Gyr, respectively. During the episode of the second infall, the inflow rate peaked at $4.13$~Gyr with a value of $20.15$~M$_{\odot}$yr$^{-1}$, and the outflow rate followed this behavior at $4.75$~Gyr with a maximum of $6.80$~M$_{\odot}$yr$^{-1}$. The outflow rate follows the inflow function, and the time of the second peak is delayed by $0.6$~Gyr.

In the last $1-2$ Gyr, the rates of the outgoing and the incoming matter declined, approaching  zero, while a weak net inflow is still observed, which is also consistent with findings in the literature. The current galactic inflow rate was investigated by for instance \citet{M12} and \citet{LH11}, and found to be $(1.1\pm0.5)$~M$_{\odot}$yr$^{-1}$, with which our study is consistent, since we obtain $1.4$~M$_{\odot}$yr$^{-1}$. We find that the present-time global outflow rate is lower than the inflow rate, $1.21$~M$_{\odot}$yr$^{-1}$.
In accordance with our model, \citet{fox_2019} derived empirical constraints on the flow rates via the UV absorption-line high-velocity clouds. Their results of $(0.53\pm0.31)$~M$_{\odot}$yr$^{-1}$ and $(0.16\pm0.10)$~M$_{\odot}$yr$^{-1}$ describing the lower limits of inflow and outflow, respectively, also suggest that the MW is currently in an inflow-dominated state. Substantial mass flux in both directions supports a galactic fountain model, in which gas is constantly recycled between the disk and the halo \citep{fox_2019}.

The global SFR throughout the simulation time is presented in the second panel of Fig.~\ref{present_v_global}. A pattern similar to the inflow rate can be seen with a delay, as prescribed by Eq.~(\ref{eq:kennicutt}). Consequently, the SFR also shows two peaks occurring after the times of the first (SFR$_{\rm max, 1}=7.15$~M$_\odot$yr$^{-1}$) and second infall (SFR$_{\rm max, 2}=7.42$~M$_\odot$yr$^{-1}$) events, separated by a gap reaching its minimum at $2.25$~Gyr. 
The second peak in the SFR occurs 4.90~Gyrs after the first one. 
Compared to the second peak of the inflow rate, the SFR shows a delay of $0.3$~Gyr. A present-day value of $1.26$~M$_\odot$yr$^{-1}$ is predicted at the end of the simulation, which lies within the observed range of $1.54 \pm 0.56$~M$_\odot$yr$^{-1}$, determined from young stellar objects detected by Spitzer \citep{RW10}. 

The occurrence rates of SNe types Ia and II during the simulation are depicted in the third panel of Fig.~\ref{present_v_global}. Type II SNe follow the trend exhibited by the inflow-driven SFR, while SNe Ia have a characteristic delay. Both curves have the two-peak shape. 
The explosion rates of SNe types II and Ia reaches a minimum within the gap at Galactic ages of $2.15$~Gyr and $3.05$~Gyr, respectively. The first and second bursts of the rate of SNe type II 
are separated by $4.4$~Gyr (exactly as the SFR), while the rate of SNe type Ia experienced 
a broader gap of about $5.9$~Gyr. In the Galactic disk, the final values reached in our model are $0.858\cdot10^{-2}$~yr$^{-1}$ for SNe~II and $4.313\cdot10^{-3}$~yr$^{-1}$ for SNe~Ia, where we applied \citet{P11} as an observational benchmark for the simulations (see Table~\ref{tab:present_values}). 
The number of SNe type~II tends to be underestimated, though this issue was also noticed by \citetalias{spitoni_2021}. We note that varying the slope of the IMF can not solve this underestimation. If we increase the upper mass limit for SNe type II, the abundance of the $\alpha$-elements and the SNe~II rate is not significantly higher. Possible extra sources, like magneto-rotational SNe, may be implemented, that would lead us beyond the scope of this paper. 

The evolution of the mass of the gas, $M_{\rm gas}$, has two peaks with a gap at $\sim 2$~Gyr. The strictly increasing function is derived by the sum of the differences between the mass locked in stars and the stellar ejecta (mass loss). These masses are plotted as functions of the Galactic age in the bottom panel of Fig.~\ref{present_v_global}. The final mass of the gas reached is $(0.84\pm 0.17)\times 10^{10}$~M$_{\odot}$ which is consistent with the observation of $(0.81 \pm 0.45)\times 10^{10}$~M$_{\odot}$ \citep{K15}. We also have good agreement with the present-day value of the total stellar mass as we obtained $(3.4 \pm 0.1)\times 10^{10}~{\rm M_{\odot}}$, near the reference value of $(3.5 \pm 0.5)\times 10^{10}~{\rm M_{\odot}}$ found by \citep{F06}.

\subsection{Evolution of the disk along the Galactocentric radius}\label{sec:along_radius}

In this section, we evaluate the regional results, and assess the consistency with the global GCE model of the MW. The regional and global best-fit model results are summarized in Table~\ref{sum_results}. The inner zone was fitted based on a total of 69,057 stars, the middle zone contains the highest number of 271,188 stars, that encompasses the solar neighborhood, while only 53,798 stars are contained in the outer regions. 

Within the six modeled regions in the top panel of Fig.~\ref{fig:res_inflow}, the time of the second infall event ranges between $t_{\rm max}=2.24$~Gyr and 4.66~Gyr. It happened earlier in the outermost regions, and as time elapsed it successively reached the regions closer to the Galactic center (see also Table~\ref{sum_results} and the bottom panel of Fig.~\ref{fig:results_times}). This suggests that the gravitational interaction between the potentially accreted dwarf galaxy and the early MW started about 11.3~Gyr ago, at the distance of $R_6=14$~kpc and finished about 2 Gyr later, 9.5~Gyr ago at $R_1=4$~kpc. The global model also suggested that it happened around the Galactic age of 4.13~Gyr, thus 10~Gyr ago, on average. We construct the sum of the masses accreted in each region consistently with the best-fit global model. As a result of this consistent coupling between the regional and global models, the global present-day inflow rate agrees well with the sum of the regional values.

In each region, there is a pause in the inflow rate between the first and second infall events at the Galactic ages of $1.25-1.85$~Gyr. The regional inflow curves in the top panel of Fig.~\ref{fig:res_inflow} show that the lowest inflow rate was usually experienced earlier in regions farther from the Galactic center. It could be also explained by a merger event affecting the outer Galaxy first. Moreover, the maximum values of the inflow rates associated with each infall event have a significant difference within the inner zone. In contrast, in the middle and outer zones, the peaking values are generally similar. The maximum inflow rate during the merger typically ranges between 0.70~M$_{\odot}$yr$^{-1}$ and 5.97~M$_{\odot}$yr$^{-1}$. The highest value corresponds to the second region centered at 6~kpc, then decreases toward the outer regions along with the shifting to earlier Galactic ages. We find that the difference between the first and second maximum values of the inflow rate is lower as the distance from the center increases. In addition to the inflow, the regional fits show that the present-day inflow rate is 0.08~M$_{\odot}$yr$^{-1}$ in the innermost region, whereas it exhibits an increasing trend and reaches a final value of 0.21~M$_{\odot}$yr$^{-1}$ in the outermost region. The sum of the six present-day values is 0.895~M$_{\odot}$yr$^{-1}$, which is consistent with that of obtained by the global fit, $0.737$~M$_{\odot}$yr$^{-1}$, and also with the references in Table~\ref{tab:present_values}.

The bottom panel of Fig.~\ref{fig:res_inflow} shows that the SFR follows a delayed trend compared to the inflow rate. It also peaks at different times as the Galactocentric distance varies. The first peak in the SFR is associated with the rapid formation phase, around 0.25~Gyr in the six annular regions and globally. This suggests  a $\approx$100-Myr-long delay from the inflow during the formation of the Milky Way. The further the region, the earlier the second SFR burst (probably triggered by the merger) is. The second SFR burst peaks at 4.25~Gyr in the sixth region and 5.25~Gyr in the first region. Compared to the time of the second infall event, the locus of the second starburst has a more significant delay than the first infall. It consistently becomes longer as the distance from the center increases, as the inner, middle, and outer zones experienced the delay between the infall and the SFR around $0.8$~Gyr, 1.1~Gyr, and 1.5~Gyr.

The time gap separating the two starbursts also shows a trend as it is shorter with increasing the distance, and the local minima of the six regional SFR functions shift to earlier ages. In addition, the maximum SFR after the second infall mostly decreases with the distance except for the second region. The results for the final SFR show a weak, decreasing trend from the value of 0.19~M$_{\odot}$yr$^{-1}$ in the first region to 0.12~M$_{\odot}$yr$^{-1}$ in the sixth region. These values are consistent with the global result, as the sum of each present-day rate is 1.01~M$_{\odot}$yr$^{-1}$ while the global fitted value is $1.11$~M$_{\odot}$yr$^{-1}$. Our results are in agreement with the reference of $1.54\pm 0.56$~M$_{\odot}$yr$^{-1}$ \citep{RW10}.

In Fig.~\ref{fig:results_times}, we plot the characteristic ascending and descending accretion timescales and time delay from the best-fit results. 
The $\tau_1$ parameter that drives the short relaxing time of the intense first infall does not show a significant trend with the Galactocentric radius, as it is roughly constant between $0.16 - 0.32$~Gyr, with a global value of 0.322 Gyr. 
The $\tau_2$ descending branch of the second infall lasts longer when moving to the outer regions. The furthest zone shows a significantly longer relaxing time of $6.0 - 9.4$~Gyr following the second infall event, compared to the closer zones where it ranges between $2.10 - 4.3$~Gyr. This suggests that the relaxation episode following the merger event was weaker and slower in the more distant regions. The global value derived for $\tau_2$ is 2.86~Gyr, representing an average along the disk.

\begin{figure*}
\centering
\includegraphics[width=.96\textwidth]{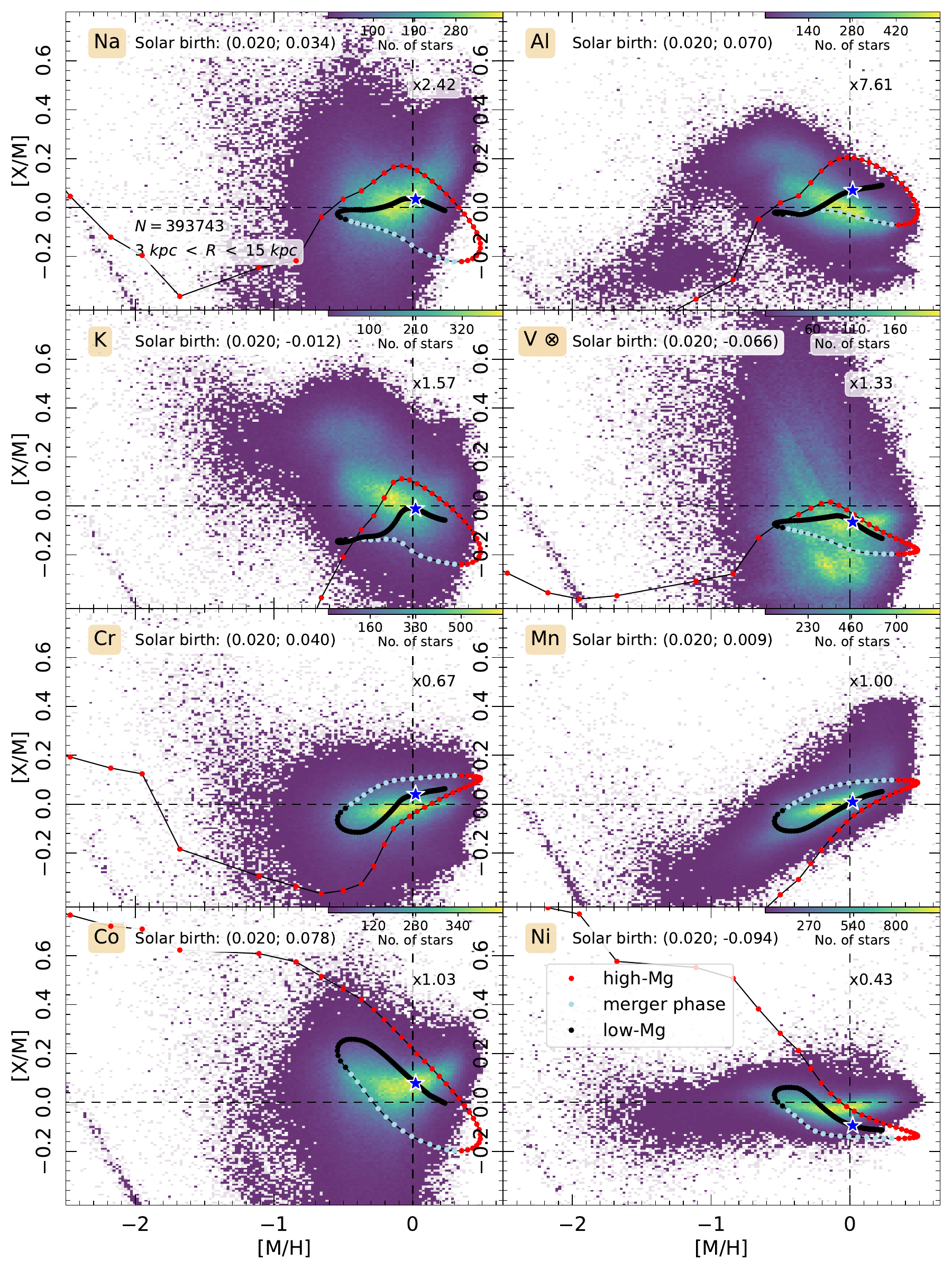}
\caption{Same as Fig.~\ref{global_fit} for the light odd-Z (Na, Al, K), iron-peak odd-Z (V, Mn, Co), and iron-peak even-Z (Cr, Ni) elemental abundances. Note that $\otimes$ drives caution to V because of high observational uncertainty.}
\label{global_fit_oddz}
\end{figure*}

\begin{table}
\centering
\caption{The applied $f_{\rm X}$ global yield multiplication factors and their discrepancies.}
\begin{tabular}{c|ccc|cc} 
\hline\hline 
  & \multicolumn{3}{c|}{our model} & \multicolumn{2}{c}{Ref. (M25)} \\
\hline  
X & $f_{\rm X}$ & $\Delta\mh$ & $\Delta{\rm [X/M]}$ &  uncertainty & {quality}\\
\hline
Mg & 3.42$\pm$0.03 & 0.020 & -0.037 & 0.053 & excellent \\
O & 2.30$\pm$0.08 & 0.020 & 0.050   & 0.056 & excellent \\ 
Si & 1.10$\pm$0.01 & 0.020 & 0.018  & 0.023 & excellent \\ 
S & 0.90$\pm$0.04 & 0.020 & 0.062   & 0.067 & good \\ 
Ca & 1.30$\pm$0.01 & 0.020 & -0.027 & 0.034 & excellent \\ 
Ti & 1.50$\pm$0.12 & 0.020 & -0.008 & 0.082 & fair \\ 
Na & 2.42$\pm$0.15 & 0.020 & 0.034  & 0.167 & fair \\ 
Al & 7.61$\pm$0.21 & 0.020 & 0.069  & 0.049 & good \\ 
K & 1.57$\pm$0.11 & 0.020 & -0.012   & 0.093 & good \\ 
V & 1.33$\pm$0.07 & 0.020 & -0.066  & 0.265 & poor $\otimes$\\ 
Cr & 0.67$\pm$0.01 & 0.020 & -0.002  & 0.078 & good \\ 
Mn & 1.00$\pm$0.01 & 0.020 & 0.009  & 0.035 & good \\ 
Co & 1.03$\pm$0.12 & 0.020 & 0.079 & 0.144 & fair \\ 
Ni & 0.43$\pm$0.02 & 0.020 & -0.024 & 0.027 & excellent \\ 
\hline
\label{tab:sum_yields}
\end{tabular}
\tablefoot{The discrepancies $\Delta$[M/H] and $\Delta$[X/M] are calculated from the solar zero-point at the time of the birth of the Sun. The last two columns contain the estimated abundance precision based on solar neighborhood and the overall assessment category of precision adopted from Meszaros et al. (2025, under rev.). Note that $\otimes$ drives caution to V. The $f_{\rm X}$ are dimensionless scaling factors, while the other columns are in units of dex.}
\end{table}

According to our six regional models, the $\tau_{\rm up}$, that represents the intensity of the exponentially ascending branch of the second infall event, varies between $0.29$ and $1.23$~Gyr throughout the regions, and is 0.55~Gyr for the best-fit global model, consistent with the regional results. This ascending phase of the accretion lasts the longest within the first region and declines until the third region centered at 8~kpc. It does not show any significant variation or trend in the outer zone. This generally suggests a more moderate accretion in the middle and outer zone compared to the inner one.

The surface mass density of the infalling matter captured during the entire Galactic evolution is $\sigma_{\rm tot}=161.5$~M$_{\odot}$pc$^{-2}$, which was distributed onto the six regions as discussed in Sect.~\ref{om_impl}, and listed in Table~\ref{sum_results}. 
The regions show an exponentially declining trend of surface mass density as a function of distance. However, the $\sigma_2 / \sigma_1$ ratio is $7.614$ in the global model, while it spans an interval of $[1.934;14.893]$ in the six separated regional models. This means that the density ratio increases toward the outer regions, while it has a value higher than one in each region. In other words, in the very first region, the second infall event provided a gas surface density two times larger than the first infall, then moving to the last region, this ratio grows until over 14 (see Fig.~\ref{fig:res_sigmas}).

These values show that although the total mass of the infalling matter decreases toward the outermost region (86.7~M$_{\odot}$yr$^{-1}$ in the inner zone, 51.0~M$_{\odot}$yr$^{-1}$ in the middle zone, and 23.7~M$_{\odot}$yr$^{-1}$ in the outer zone), the mass proportion originating from the second infall gets more significant. 
Therefore, by moving toward the outer regions their mass are more and more dominantly increased by the merger event. 
The results obtained here are consistent with those of \citetalias{spitoni_2021}, where the GCE models having a resolution of three regions suggest the following. The inner galactocentric zone enclosed between 2 and 6~kpc received surface densities with a ratio of $3.805^{+0.078}_{-0.113}$, the middle zone between 6 and 10~kpc is best fit with a ratio of $5.635^{+0.214}_{-0.162}$, and the outer one defined from 10 to 14~kpc is best fit with a ratio of $10.348^{+0.188}_{-0.171}$.

The fact that the inner Galactic regions have shorter infall timescales in the high-Mg phase suggests a faster formation and assembly compared to the outer regions, supporting the inside-out formation scenario. As demonstrated by the observational data in Fig.~\ref{regs_only_mg}, the locus of the low-Mg sequence is shifted toward lower metallicities as the distance increases from the Galactic center. These outer regions may experience a weaker and less efficient chemical enrichment as a result of longer accretion timescales. Therefore, a generally lower metallicity and Mg-abundance was reached at the end of the model  (Fig.~\ref{fig:feh_vs_age_regional}). 
In addition, the high-Mg phase has fewer stars as a function of Galactocentric distance, and we associate the more prominent low-Mg group with a larger surface density ratio there. A similar trend of growing $\sigma_2 / \sigma_1$ ratio was found by \citet{palla_2020}.

\subsection{Evolution of metallicity and magnesium}

Figures~\ref{fig:feh_vs_age} and \ref{fig:feh_vs_age_regional} show the global and regional evolution of the [M/H], [Mg/H], and [Mg/M] abundance ratios in the stars as a function of Galactic age, where vertical lines mark the times of the local maximum and minimum points from the simulation. 
The [Mg/H] and metallicity display similar evolutionary tracks as a function of time, while the [Mg/M] shows an inverse trend. The [M/H] versus [Mg/M] are plotted in the first panel of Fig.~\ref{global_fit}.

As it can be inferred from Fig.~\ref{fig:feh_vs_age} (black model curve), the [Mg/M] abundance reaches its highest value shortly after the formation period, and when SNe~Ia start to contribute, [Mg/M] generally declines between 1-3~Gyr, while both metallicity (blue curve) and [Mg/H] (red curve) are sharply rising until $t\approx2$~Gyr. According to our best-fit model covering the entire disk, magnesium and metals decline relative to hydrogen after $t\approx 2$~Gyr, when the second infall event begins.

Subsequently, this merger event causes an enrichment of Mg relative to metallicity at around 4~Gyrs (see the rising phase of the loop in the first panel of Fig.~\ref{global_fit}). During this period, it significantly increases the rate of SNe~II, while the SNe~Ia reach their maximum with a delay of $\sim$2~Gyr (see the third panel of Fig.~\ref{present_v_global}). The growing number of exploding white dwarfs again causes a moderately steep decline of [Mg/M]. Shortly after the peak of the second infall at $t_{\rm max}$, there is a minimum of Mg and metallicity relative to H, while the [Mg/M] had a temporary maximum. Afterward, the [Mg/H] and metallicity steadily increase, while [Mg/M] mildly decreases for about 9 Gyr to the present time. Throughout the entire simulation, [M/H] and [Mg/H] show similar trends, in a way that the overall metallicity [M/H] is delayed by $<$0.5~Gyr from [Mg/H].

The abundance ratios of [M/H], [Mg/H], and [Mg/M] in the six regions are depicted in Fig.~\ref{fig:feh_vs_age_regional}. When comparing their evolution curves, we see that the minimum [Mg/M] reached after the formation phase tends to be consistently larger at regions at greater Galactocentric distance, and occur progressively earlier in the model (at time from 2.4 to 3.2~Gyr). Then at around 2.7-4.5 Gyr, due to the burst in SNe type II induced by the merger event, [Mg/M] experiences a local maximum, and monotonously declines until the present day as SNe~Ia occur at a longer and delayed timescale. The maximum [Mg/M] is reached at $\sim 6$~Gyr in the R$_1$ innermost region and $\sim$3.1 Gyr in the R$_6$ outermost region.

The [M/H] and [Mg/M] reach a temporary maximum between the first infall event's relaxation phase and the second infall's ascending branch. These maxima happened the earliest at a Galactocentric distance of 14~kpc ($\sim$1.5~Gyr) and later at $\sim$2.2~Gyr within the region centered at 4~kpc. The second peak of the infall rate causes a sudden drop in both abundances, which is seen as a loop in, for example, Fig.~\ref{global_fit}. These minimum values become larger within the inner zone, while the Galactic age is almost 5~Gyr in the first region and $\sim$3~Gyr in the sixth one. The present-day abundance values can also be observed in the regional panels of Fig.~\ref{fig:feh_vs_age_regional} as the final values of the evolutionary curves. Our model predictions are consistent with the observations, which predict a declining present-day radial abundance gradient of [Mg/H] and [M/H]. These trends in the regional results support the presence of a major merger event in Galactic history that first started to interact with the outer MW and then gradually approached the inner Galaxy.

\subsection{Evolution of the individual elements}

\subsubsection{Evolution versus metallicity}

The best-fit evolutionary curves for the elements are shown in Figs.~\ref{global_fit} and \ref{global_fit_oddz}, compared to the SDSS-V MWM DR19 observations. The reversed feature in the [M/H] evolution generally appears around the Galactic age of 4~Gyr when the intense accretion of primordial-composition gas occurs, diluting the ISM and thus reducing the overall metallicity. The evolution of each $\alpha$-element as a function of metallicity is depicted in Fig.~\ref{global_fit} for the entire disk. 
The observed chemical evolution patterns of Mg, Si, S, and Ca are reproduced. On the other hand, for O and Ti GCE predictions also tend to disagree during the merger event and the low-Mg phase, independently from any correction factor applied to the theoretical stellar yields. 
The observed abundances of these latter elements are derived with a typical uncertainty of 0.056~dex and 0.082~dex in the solar neighborhood sample, respectively (M25). Oxygen is determined with an excellent quality, involving a slight zero-point offset, no significant dependency on effective temperature, high precision (M25), and therefore, it can be used safely throughout a large enough parameter space covered by our study (as detailed in the target selection in Sect.~\ref{sec:target}). The model behavior at the low metallicity of [M/H]$\approx$-2~dex exhibits an under estimation of abundances of S and Ca, while an over estimation of Ti compared to the observation. 

The thin and thick disks are sharply separated on the chemical map drawn by the observations of O as well. However, it has been noted before that there are difficulties in deriving [O/M] from the atomic absorption lines in optical wavelength ranges \citep[e.g., GALAH,][]{desilva_2015} and from the molecular lines in near-infrared spectra of stellar atmospheres (APOGEE). As studied on golden cross-matched samples and also outlined by \citet{hegedus_2023}, the discrepancies between the [O/H] abundances derived by sky survey programs range from $0.03$ to $0.20$~dex for giant stars. Therefore, one observes disagreements in the slopes displayed by APOGEE and GALAH data of the [M/H] versus [Mg/O] high-$\alpha$ and low-$\alpha$ sequences \citep[see, e.g., ][]{weinberg_2019, griffith_2019}. Moreover, the median $\Delta$[O/H] abundance difference between MWM DR19 and GALAH DR3 is $-0.05$~dex with a significant scatter of $0.22$~dex (M25). In contrast, titanium is the least precise $\alpha$-element published in MWM DR19 and was derived with fair precision and accuracy due to the fact that Ti--with its relatively weak absorption lines--was possible to be measured reliably only in a narrow parameter region, between 4000~K~$\leq T_{\rm eff}\leq$~5500~K and around solar metallicity for giants. While the median $\Delta$[Ti/H]~$=-0.05\pm0.28$~dex difference and scatter as compared to GALAH DR3 is similar to the case of O, the significant temperature dependence of [Ti/H] detection in the solar neighborhood may be complex. The discrepancies might be due to the difficulty of fitting the mostly blended Ti lines or NLTE effects (M25).

The evolution of the light odd atomic number and iron-peak  elements (Na, Al, K, V, Cr, Mn, Co, and Ni) is depicted in Fig.~\ref{global_fit_oddz}. The abundances of Cr, Mn, and Ni are well tracked throughout the chemical maps, whereas the distribution patterns of Na, Al, K, and Co are only fairly reproduced. In most cases, especially of K, Cr, and Ni, we still observe a discrepancy at low metallicity, whereas obtaining a good agreement near the solar value at the same time. Generally, we modeled the iron-peak elements more accurately than the light odd-Z elements. Among the iron-peak elements, the MWM Ni abundances have excellent quality, and can be used reliably above 4250~K, according to the estimated overall precision. Moreover, there is a very good agreement with GALAH DR3 (M25). On the contrary, V abundances are categorized as of poor quality, and it is advised to use [V/H] values with caution (M25). The comparison detailed in M25 with GALAH DR3 data showed a strong correlation with $T_{\rm eff}$ and a mean discrepancy of $-0.13\pm0.43$~dex. The other elements in Fig.~\ref{global_fit_oddz} are either determined with a good or fair accuracy and precision in MWM. The reliable parameter ranges for these elements are typically restricted (M25). The reason for the discrepancies between the model and the MWM data may be a result of both observational uncertainties (e.g., degeneracies between age and [M/H] \citep{jofre_2019}, LTE approximations, abundance derivation systematics) and model uncertainties (e.g., reaction rate uncertainties, space progenitor grids, and GCE model assumptions).

\subsubsection{Evolution versus magnesium}

\begin{figure*}
\centering
\includegraphics[width=.98\textwidth]{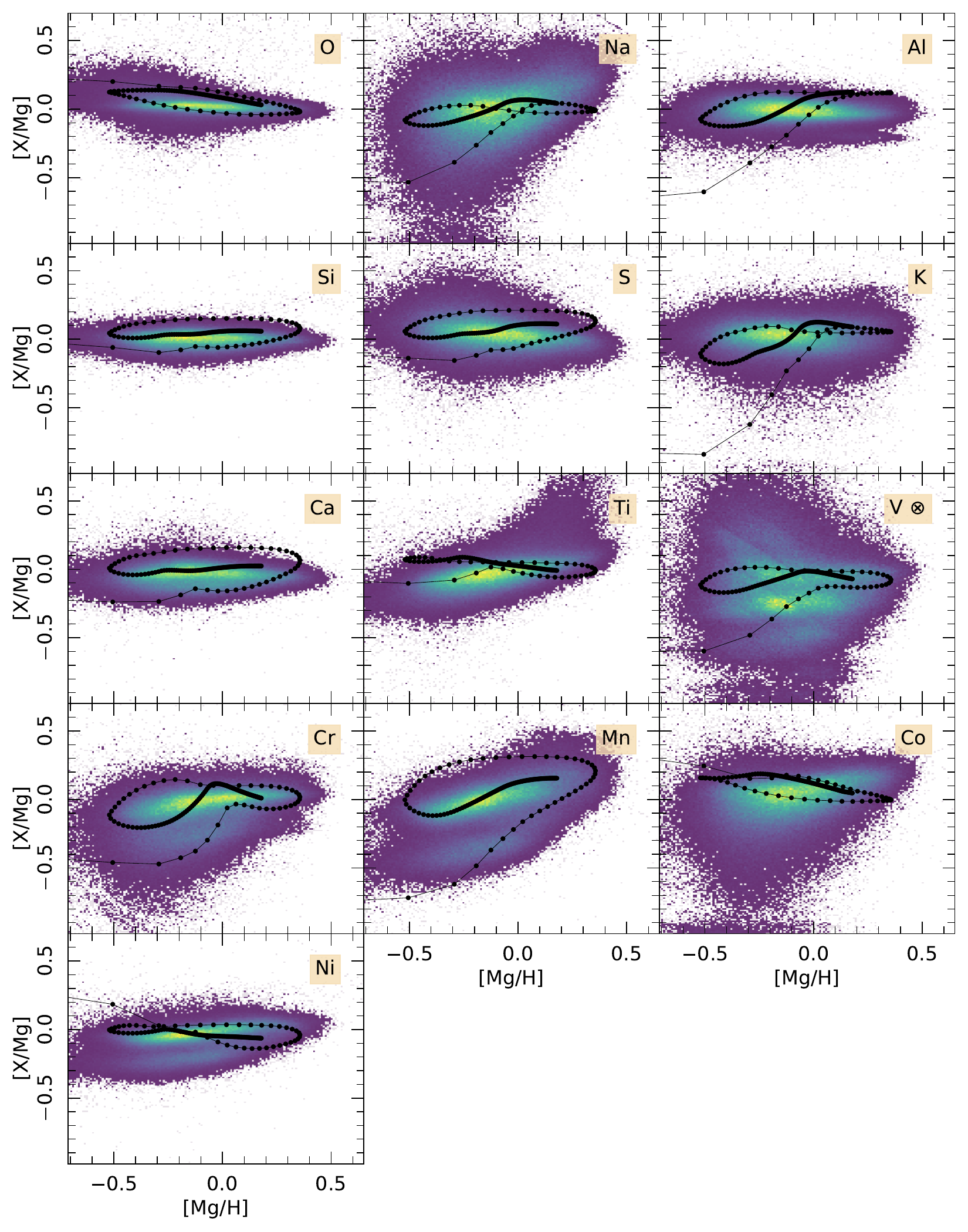}
\caption{ Observed [X/Mg] vs. [Mg/H] abundance ratios for the $\alpha$- and odd-Z elements from MWM DR19 throughout the entire galactocentric region between 3 and 15~kpc compared with the best-fit CE model results (dotted curves) for all species. The color coding represents the number of stars on the observational chemical map of the MW, and gray squares represent bins containing a single star. Note that $\otimes$ drives caution to V because of high observational uncertainty.}
\label{fig:mg_reference}
\end{figure*}

Figure~\ref{fig:mg_reference} shows the [X/Mg] abundances as a function of [Mg/H]. For the elements produced predominantly by CCSNe, the observations are expected to exhibit a uniform horizontal trend in the [X/Mg]--[Mg/H] parameter plane. For instance, in the first panel of Fig.~\ref{fig:mg_reference}, the O abundances from the MWM data show this trend with a relatively small scatter. The evolutionary curve has an $\approx$0.1~dex discrepancy, which is the result of not being able to fit both Mg and O accurately (see the first row of Fig.~\ref{global_fit}). 

Generally O and Al are produced by CCSNe like Mg, thus exhibiting a horizontal evolution along with the [Mg/H] abundance. The observation and theory normally agree, except, the distribution of the measured Na abundances, which steadily increases with [Mg/H]. We also see that the model for Al and K shows a slight [Mg/H] dependency not seen in the data. Na has only two, reasonably weak lines in the H-band and might be also affected by minor errors introduced by the telluric correction. For Al, a possible difficulty  could be NLTE effects, which were not taken into account in the spectral grid in DR19 (M25). Si and K, which are dominantly produced by CCSNe, show divergent results. Silicon has a good agreement, while the model for [K/Mg] shows disagreements at [Mg/H]~$\lesssim 0$ and an offset from the observation. 
The S and Ca are produced in both CCSNe and SNe type Ia. Our GCE model for S reflects a minor discrepancy. Our model and the Ca measurements have a very good agreement. 

Other elements such as Ti, V, Cr, Mn, Co, and Ni have contributions dominantly or closely exclusively from SNe~Ia. Our model for Mn well fits the observational thin and thick disk sequences. Titanium and cobalt display a fair agreement with slight offsets of $\sim$0.1~dex, while Cr fits only moderately the observations with a significant offset discrepancy. We would recommend to not use V abundances, since there is no recommended reliable parameter range in M25.  

To summarize, the elements that are difficult to measure observationally, due to the lack of clearly detectable or well-defined spectral lines, have fair or poor quality values published in DR19. This is the case for Na, Ti, Co and V (M25), therefore, we cannot derive strong conclusions yet from their comparison with GCE models. On the contrary, the observed abundances of O, Si, Ni, Al, S, K, Cr, and Mn are reported to have excellent or good reliability in general (M25). There still may be some unrevealed issues for Al, K, and Cr, though, such as the lack of NLTE corrections or unknown line blendings, as these elements exhibit discrepancies between the GCE model and MWM spectroscopy.

\subsubsection{Yield modifications}

The $f_{\rm X}$ yield multiplication factors are found during the global fit and then kept fixed when completing the regional fits (Table~\ref{tab:sum_yields}). 
As explained in Sect.~\ref{sec:yields}, $f_{\rm X}$  account for discrepancies between the observations and the theoretical yields. If the yields are correct and can reproduce the observed abundance trends, then the $f_{\rm X}$ values should be close to one. However, we find that $f_{\rm X}$ needs to be greater than one to fit the Mg, O, Na, Al, and Co abundances, while we applied $f_{\rm X}$ less than one for Cr and Ni. The best fit yield correction factors show that the theoretical yield sets underlying our models systematically overproduce Cr and Ni and underproduce Mg, O, Na, Al, and Co. Such disparities between theoretical yields and observations exist for most yield sets \citep[e.g., ][for CCSN yields]{griffith_2021}. If a different theoretical yield set is used, the factors found in this work will be varied, while the parameters describing the inflow function will practically provide an unaltered GCE picture of the MW. Most of the theoretical yield noise is captured by $f_{\rm X}$, without affecting the relevant results for our study, but as this factor is not metallicity dependent, it cannot fully correct for the yield errors.  

We assume that stars formed at the same time as the Sun have the same solar chemical composition. This assumption is well established within the solar neighborhood, although it becomes an approximation when applied to the entire Galactic disk. To fulfill this assumption, the (\ref{eq:abund_def}) definition of elemental abundance ratio [X/M] specifies that the individual values for the 14 chemical abundances have to be close to [M/H]=[X/M]=0 for an SSP formed together with the Sun in time. We found a metallicity residual of $\Delta$[M/H]=0.020~dex, and between $-0.066$ and $0.069$~dex for the $\Delta{\rm [X/M]}$ individual abundances. 
(See Table~\ref{tab:sum_yields} and Figs.~\ref{global_fit}, \ref{global_fit_oddz}.) 
Considering the typical uncertainties of the abundance determination from the stellar spectra (M25), these residuals are still in good agreement both with the observations and theoretical predictions. According to the precision classification terminology for DR19 abundances in M25 (detailed in their Table~5), Mg, O, Si, Ca, Mn, and Ni have an excellent precision of lower than 0.1~dex, while S, Al, K, and Cr have a good precision between 0.10 and 0.15~dex, and Ti, Na, V and Co exhibit a fair or poor precision of above 0.3~dex. Note that the uncertainties provided here did not include accuracy and systematics. The abundance residuals are generally below or have the same order as the reported precision levels.

\subsection{Comparison to \citet{spitoni_2021}}\label{comparison}

The simulations presented in this work are based on a revised two-infall scenario and support an inside-out formation as \citetalias{spitoni_2021}. 
Their comparison target selection ranged between the Galactocentric distances of 2~kpc and 14~kpc, separated into three regions. These regions contained 26,690 MW-disk stars from the APOGEE DR16 data set \citep{ahumada_2020} and were used to fit the model. Here we use an expanded data set including 393,743 stars with a higher spatial resolution separated into six annular regions between 3--15~kpc. 
For comparison, the results for the inflow timescales, the time of the second infall, and the ratio of the surface mass densities obtained by \citetalias{spitoni_2021} are listed in Table~\ref{sum_results} (global model) and in Figs.~\ref{fig:results_times} and \ref{fig:res_sigmas} (regional models).

The inflow function directly before the second peak is constant at zero in \citetalias{spitoni_2021}, while our model applies a rising branch of accretion as well. 
Both models neglect the effect of radial stellar migration in the disk. Effective net gas outflows were not introduced by \citetalias{spitoni_2021}, whereas they invoked a previously enriched gas infall to correctly reproduce the [Mg/Fe] vs. [Fe/H] abundance ratio of the low-$\alpha$ sequence in the inner disk. Our GCE model involves a constant mass-loading parameter to drive the galactic winds permanently leaving the Galactic disk while allowing for an infall with primordial composition matter.

Both in the present work and in \citetalias{spitoni_2021}, Mg and Fe were used as key tracers, while the chemical maps of 14 elements beyond [Mg/Fe] vs. [Fe/H] were also reproduced here. As the source for the nucleosynthetic yields, \citetalias{spitoni_2021} adopted the yield set of \citet{ww_95} for massive stars, and \citet{iwamoto_1999} for SNe type Ia. In their best-fit model, Mg and Si yields were artificially raised. In contrast, our study is based on the theoretical yields composed and published by the NuGrid collaboration and by FRUITY \citep{fruity} for massive and AGB stars, respectively. The yields of SNe Ia come from the same reference work. We keep the yields of Fe untouched, also $f_{\rm Fe}=1$, as in \citetalias{spitoni_2021}, while the individual elemental contributions were modified according to Table~\ref{tab:sum_yields}.

As shown in Table~\ref{sum_results} (for the global model) and Fig.~\ref{fig:results_times} (for the regional models), the values and trends describing the time of the merger event and the accretion timescales generally agree well with \citetalias{spitoni_2021}. The agreement is also excellent in the region where the Sun is located (Fig.~\ref{fig:results_times}).
The global and regional fit $\sigma_2 / \sigma_1$ are listed in Table~\ref{sum_results}, and plotted in Fig.~\ref{fig:res_sigmas}. Although their number of regions and the spatial boundaries of the separations differ from those we used, an increasing trend is also observed with Galactocentric distance.

Minor differences between our results and \citetalias{spitoni_2021} originate from using different observational data sets, not using the exact identical chemical model assumptions and computational techniques, and adopting different nucleosynthetic yield sets. In general, both results reflect a similar picture of Galactic history: our model and \citetalias{spitoni_2021} share a similar growth of the Galactic disk following the inside-out scenario and predict analogous metallicity distribution functions. The potential merger event at a Galactic age of around 4 Gyr, causing a well-defined gap in the SFR, is confirmed as well.

\section{Conclusions}\label{conclusions}

We study the GCE of 14 elements based on a golden sample of 393,743 stars from the first data release (DR19) of the MWM spectroscopic sky survey part of SDSS-V. First, we analyze the observed chemical map of our Galaxy. The separation between the chemical thin and thick disks is defined using [Mg/M], and we discuss the high- and low-Mg sequences across the Galactocentric radii. The chemical maps of most elements considered here show similar trends to Mg, although sequences of odd-Z species such as Co, Mn or V show different patterns as functions of metallicity. Following previous studies \citep[e.g.,][]{hayden_2015}, but with an order of magnitude more stars, we find that the locus of the low-Mg sequence is gradually shifted toward lower metallicity in the outer regions, while the number of stars in the high-Mg population decreases both absolutely and relatively to the low-Mg group. 

We present extended chemical evolution models for the Galactic disk and six galactocentric regions between $3~\kpc$ and $15~\kpc$. The measured abundances of Mg constrained the model by utilizing the semi-analytical GCE code \texttt{OMEGA+}. We consider seven free global parameters related to the two infall episodes: the accretion relaxing time-scales $\tau_1$ and $\tau_2$, the characteristic duration of the merger phase $\tau_{\rm up}$, the delay $t_{\rm max}$ and the ratio and sum of the surface mass densities associated to the second and first infall events $\sigma_2/\sigma_1$ and $\sigma_{\rm tot}$, and the Mg yield scaling factor $f_{\rm Mg}$. While performing the regional fits, the variables $\sigma_{\rm tot}$ and $f_{\rm Mg}$ are kept fixed. The following conclusions describing the formation history are made by fitting the observational chemical maps:

\begin{enumerate}
        \item The here obtained \texttt{OMEGA+} model suggests a general agreement with the recent MWM observations published in DR19 (M25). A significant delay time between the two gas infall episodes for the thick-disk and thin-disk formation is confirmed in all analyzed galactocentric regions. We find that the approximate value for the delay time ranges between 2.6 and 4.7~Gyr, confirming the results of \citetalias{spitoni_2021}, but on a much larger sample. Therefore, we confirm the concept of a merger event approximately $10$~Gyr ago in the Galactic history.
        
        \item This prolonged merger event starts around $3$~Gyr after the initial formation of the MW in the outer zone and $5$~Gyr in the inner zone. 
        As the Galactocentric radius is growing, a decrease in $t_{\rm max}$ produces a tighter loop starting at lower metallicities and allowing for only a higher value for the minimum of the Mg abundance during the infall event \citep{spitoni_2019}
        When the Galaxy reaches the age of $8$~Gyr, the rate of accreting matter dramatically falls (Fig. \ref{fig:res_inflow}). Since our model represents  a two-infall scenario, the first inflow event creates the high-$\alpha$ (or high-Mg) thick disk stars during Galactic formation, and the second allows for the birth of the low-$\alpha$ (or low-Mg) stars of the thin disk (Fig. \ref{global_fit}). 
        
        \item The entire merger event lasts for about $5$~Gyr and occurs earlier at larger Galactocentric distances. The ratios between the thin and thick disk surface mass densities grow toward the outer regions, in agreement that as we move toward external regions, the distribution of the MWM DR19 data sample represents fewer stars in the high-Mg phase compared to the low-Mg sequence. We derive an inside-out formation of the thin-disk of our Galaxy according to the best fit of our multi-zone CE simulation, meaning that the inner Galactic regions are assembled on a shorter timescale than the external parts.

        The beginning of the merger may be estimated with the characteristic rising time of the accretion interval as $t_{\rm max} - \tau_{\rm up}$, therefore our results confirm the conclusions of \citet{chaplin_2020}, as we find that the earliest time the merger could have begun was 11.5$-$11.9~Gyr ago in the outermost region.
        
        \item Important observational constraints like the present-day inflow rate, star formation rate, SNe rates, and the stellar and gas mass are reasonably well reproduced within the uncertainties of observations. The six regional results are consistent with the global best-fit model for obtaining the present-day global inflow rate/SFR/stellar and gas mass by the sum of the regional values. 
        
        \item The chemical evolution is substantially driven by the stellar yields. The solar composition is generally obtained within observational uncertainties (M25). The evolutionary curves show good agreement for the $\alpha$-elements Mg, Si, Ca, and for the odd-Z elements Na, Mn, Ni. Elements Al, K, Ti, and V are not well reproduced. Discrepancies may originate from observational and model uncertainties, including unresolved or heavy blending in the spectra, LTE approximations in the abundance determination, uncertainties in theoretical reaction rates, and assumptions about massive star black hole formation. Our implementation of yield correction factors allow for the relevant GCE results to be obtained relatively independently of the chosen yield set, with a $\lesssim$10\% variation. 
        Moreover, when performing the optimization based on Si instead of Mg, the GCE results  suggest the same formation history with an inflow function characterized by a time of the second peak, inflow timescales, and ratio and sum of the surface mass densities, that only changes by $\lesssim$15\%. 
\end{enumerate}

We generally confirm the results obtained by \citetalias{spitoni_2021}, adopting similar GCE parametrization but using different stellar yields and a different set of observations to benchmark the GCE calculations.  
Our study modeling the merger/dual-infall Galactic CE for the disk also represents a complementary scenario as that proposed by \citet{sharma_2020} and \citet{pranztos_2023}, in which the radial variation in the [$\alpha$/M] vs. [M/H] abundance parameter plane has been explained with stellar migration. 
In the future, the implementation of stellar migration, the variation of the SFR along with the Galactocentric distance, and the use of the subsequent most recent data sets from SDSS-V will serve as a development to this framework.

\begin{acknowledgements}

VH and SzM acknowledge the support of Hungarian Academy of Sciences through MTA-ELTE Lend{\"u}let ``Momentum'' Milky Way Research Group. 
Supported by the DKOP-23 Doctoral Excellence Program of the Ministry for Culture and Innovation from the Source of the National Research, Development and Innovation Fund and Lendület LP2023-10. 

BV, MP, and ML acknowledge the support of the Hungarian Academy of
Sciences via the Lend\"ulet Program LP2023-10. This work was supported
by the European Union’s Horizon 2020 research and innovation programme
(ChETEC-INFRA -- Project no. 101008324), and the IReNA network
supported by US NSF AccelNet (Grant No. OISE-1927130). ML was also
supported by the NKFIH excellence grant TKP2021-NKTA-64.

E.J.G. is supported by an NSF Astronomy and Astrophysics Postdoctoral Fellowship under award AST-2202135.

D.S. thank the National Council for Scientific and Technological Development – CNPq process No. 404056/2021-0.

Funding for the Sloan Digital Sky Survey V has been provided by the Alfred P. Sloan Foundation, the Heising-Simons Foundation, the National Science Foundation, and the Participating Institutions. SDSS acknowledges support and resources from the Center for High-Performance Computing at the University of Utah. SDSS telescopes are located at Apache Point Observatory, funded by the Astrophysical Research Consortium and operated by New Mexico State University, and at Las Campanas Observatory, operated by the Carnegie Institution for Science. The SDSS website is \url{www.sdss.org}.
SDSS is managed by the Astrophysical Research Consortium for the Participating Institutions of the SDSS Collaboration, including the Carnegie Institution for Science, Chilean National Time Allocation Committee (CNTAC) ratified researchers, Caltech, the Gotham Participation Group, Harvard University, Heidelberg University, The Flatiron Institute, The Johns Hopkins University, L'Ecole polytechnique f\'{e}d\'{e}rale de Lausanne (EPFL), Leibniz-Institut f\"{u}r Astrophysik Potsdam (AIP), Max-Planck-Institut f\"{u}r Astronomie (MPIA Heidelberg), Max-Planck-Institut f\"{u}r Extraterrestrische Physik (MPE), Nanjing University, National Astronomical Observatories of China (NAOC), New Mexico State University, The Ohio State University, Pennsylvania State University, Smithsonian Astrophysical Observatory, Space Telescope Science Institute (STScI), the Stellar Astrophysics Participation Group, Universidad Nacional Aut\'{o}noma de M\'{e}xico, University of Arizona, University of Colorado Boulder, University of Illinois at Urbana-Champaign, University of Toronto, University of Utah, University of Virginia, Yale University, and Yunnan University.

{For the \texttt{lmfit} Python package, Matthew Newville wrote the original version and maintains the project: https://lmfit.github.io/lmfit-py/index.html.}
{This work made use of Astropy (\url{http://www.astropy.org}): a community-developed core Python package and an ecosystem of tools and resources for astronomy \citep{astropy:2013, astropy:2018, astropy:2022}.}

{This research was supported by a grant from the European Astronomical Society thanks to the generous support of the MERAC Foundation and Springer Verlag.}
\end{acknowledgements}

\bibliographystyle{aa}
\bibliography{references} 
\end{document}